\RequirePackage{fix-cm}

\documentclass[twocolumn,epjc3]{svjour3}  
\smartqed  
\RequirePackage{graphicx}
\graphicspath{{./}{./images/}}

\RequirePackage{mathptmx}      
 \usepackage{xcolor}
\RequirePackage[numbers,sort&compress]{natbib}
\RequirePackage[colorlinks,citecolor=blue,urlcolor=blue,linkcolor=blue]{hyperref}
\usepackage{amsmath} 
\usepackage{amssymb} 
\usepackage{paralist} 

\RequirePackage[separate-uncertainty=true,list-final-separator={, and }]{siunitx}
\usepackage[utf8]{inputenc} 
\usepackage{mathtools} 
\usepackage[version=4]{mhchem} 

\newcommand{\nubar}{\overline{\nu}}

\journalname{Eur. Phys. J. C}

\begin{document}

\sloppy

\title{Search for Signatures of Sterile Neutrinos with Double Chooz}



\authorrunning{The Double Chooz Collaboration (status: June 20, 2020} 

%
%
\newcommand{\Aachen}{III. Physikalisches Institut, RWTH Aachen University, 52056 Aachen, Germany}
\newcommand{\Alabama}{Department of Physics and Astronomy, University of Alabama, Tuscaloosa, Alabama 35487, USA}
\newcommand{\Argonne}{Argonne National Laboratory, Argonne, Illinois 60439, USA}
\newcommand{\APC}{APC, Universit\'e de Paris, CNRS, Astroparticule et Cosmologie, F-75006, Paris} 
\newcommand{\CBPF}{Centro Brasileiro de Pesquisas F\'{i}sicas, Rio de Janeiro, RJ, 22290-180, Brazil}
\newcommand{\CENBG}{Universit\'e de Bordeaux, CNRS/IN2P3, CENBG, F-33175 Gradignan, France}
\newcommand{\Chicago}{The Enrico Fermi Institute, The University of Chicago, Chicago, Illinois 60637, USA}
\newcommand{\CIEMAT}{Centro de Investigaciones Energ\'{e}ticas, Medioambientales y Tecnol\'{o}gicas, CIEMAT, 28040, Madrid, Spain}
\newcommand{\Drexel}{Department of Physics, Drexel University, Philadelphia, Pennsylvania 19104, USA}
\newcommand{\INR}{Institute of Nuclear Research of the Russian Academy of Sciences, Moscow 117312, Russia}
\newcommand{\CEA}{IRFU, CEA, Universit\'{e} Paris-Saclay, 91191 Gif-sur-Yvette, France} 
\newcommand{\Kitasato}{Department of Physics, Kitasato University, Sagamihara, 252-0373, Japan}
\newcommand{\Kobe}{Department of Physics, Kobe University, Kobe, 657-8501, Japan}
\newcommand{\Kurchatov}{NRC Kurchatov Institute, 123182 Moscow, Russia}
\newcommand{\MaxPlanck}{Max-Planck-Institut f\"{u}r Kernphysik, 69117 Heidelberg, Germany}
\newcommand{\NotreDame}{University of Notre Dame, Notre Dame, Indiana 46556, USA}
\newcommand{\IPHC}{IPHC, CNRS/IN2P3, Universit\'{e} de Strasbourg, 67037 Strasbourg, France}
\newcommand{\SUBATECH}{SUBATECH, CNRS/IN2P3, Universit\'{e} de Nantes, IMT-Atlantique, 44307 Nantes, France}
\newcommand{\TohokuUni}{Research Center for Neutrino Science, Tohoku University, Sendai 980-8578, Japan}
\newcommand{\TokyoInst}{Department of Physics, Tokyo Institute of Technology, Tokyo, 152-8551, Japan }
\newcommand{\TokyoMet}{Department of Physics, Tokyo Metropolitan University, Tokyo, 192-0397, Japan}
\newcommand{\Muenchen}{Physik Department, Technische Universit\"{a}t M\"{u}nchen, 85748 Garching, Germany}
\newcommand{\Tubingen}{Kepler Center for Astro and Particle Physics, Universit\"{a}t T\"{u}bingen, 72076 T\"{u}bingen, Germany}
\newcommand{\UFABC}{Universidade Federal do ABC, UFABC, Santo Andr\'{e}, SP, 09210-580, Brazil}
\newcommand{\UNICAMP}{Universidade Estadual de Campinas-UNICAMP, Campinas, SP, 13083-970, Brazil}
\newcommand{\vtech}{Center for Neutrino Physics, Virginia Tech, Blacksburg, Virginia 24061, USA}
\newcommand{\Chooz}{{LNCA Underground Laboratory, IN2P3/CNRS - CEA, Chooz, France}}
%
%
\newcommand{\SUSSEX}{Department of Physics and Astronomy, University of Sussex, Falmer, Brighton, United Kingdom}
\newcommand{\ORSAY}{IJC Laboratory, CNRS/IN2P3, Universit\'{e} Paris-Saclay, Orsay, France}
\newcommand{\Londrina}{Universidade Estadual de Londrina, 86057-970 Londrina, Brazil}
\newcommand{\MIT}{Massachusetts Institute of Technology, Cambridge, Massachusetts 02139, USA}
\newcommand{\TokyoSci}{Tokyo University of Science, Noda, Chiba, Japan}
\newcommand{\Hawaii}{Physics \& Astronomy Department, University of Hawaii at Manoa, Honolulu, Hawaii, USA}
\newcommand{\KEK}{High Energy Accelerator Research Organization (KEK), Tsukuba, Ibaraki, Japan}
\newcommand{\Arcadia}{Physics Department, Arcadia University, Glenside, PA 19038}
\newcommand{\LAPP}{LAPP, CNRS/IN2P3 , 74940 Annecy-le-Vieux, France} 
\newcommand{\IFIC}{Instituto de F\'{i}sica Corpuscular, IFIC (CSIC/UV), 46980 Paterna, Spain}
\newcommand{\SK}{Kamioka Observatory, ICRR, University of Tokyo, Kamioka, Gifu 506-1205, Japan}
\newcommand{\SD}{South Dakota School of Mines \& Technology, 501 E. Saint Joseph St. Rapid City, SD 57701}
\newcommand{\StonyBrooks}{State University of New York at Stony Brook, Stony Brook, NY, 11755, USA}
\newcommand{\UCB}{University of California, Department of Physics, Berkeley, CA 94720-7300, USA and Lawrence Berkeley National Laboratory, Berkeley, CA 94720-8153, USA}


\author{ {T.~Abrah\~{a}o}\thanksref{e,d}
\and {H.~Almazan}\thanksref{o}
\and {J.C.~dos Anjos}\thanksref{e}
\and {S.~Appel}\thanksref{v}
\and {{J.C.~Barriere}}\thanksref{k}
\and {I.~Bekman}\thanksref{a}
\and {T.J.C.~Bezerra}\thanksref{r,1}
\and {L.~Bezrukov}\thanksref{j}
\and {E.~Blucher}\thanksref{g}
\and {T.~Brugi\`{e}re}\thanksref{q}
\and {C.~Buck}\thanksref{o}
\and {J.~Busenitz}\thanksref{b}
\and {A.~Cabrera}\thanksref{d,aa,2}
\and {M.~Cerrada}\thanksref{h}
\and {E.~Chauveau}\thanksref{f}
\and {P.~Chimenti}\thanksref{e,3}
\and {{O.~Corpace}}\thanksref{k}
\and {J.V.~Dawson}\thanksref{d}
\and {Z.~Djurcic}\thanksref{c}
\and {A.~Etenko}\thanksref{n}
\and {H.~Furuta}\thanksref{s}
\and {I.~Gil-Botella}\thanksref{h}
\and {{A.~Givaudan}}\thanksref{d}
\and {{H.~Gomez}}\thanksref{d,k}
\and {L.F.G.~Gonzalez}\thanksref{y}
\and {M.C.~Goodman}\thanksref{c}
\and {T.~Hara}\thanksref{m}
\and {{J.~Haser}}\thanksref{o}
\and {D.~Hellwig}\thanksref{a}
\and {L.~Heuermann}\thanksref{a}
\and {A.~Hourlier}\thanksref{d,4}
\and {M.~Ishitsuka}\thanksref{t,5}
\and {J.~Jochum}\thanksref{w}
\and {C.~Jollet}\thanksref{f}
\and {K.~Kale}\thanksref{f,q}
\and {M.~Kaneda}\thanksref{t}
\and {{M.~Karakac}}\thanksref{d}
\and {T.~Kawasaki}\thanksref{l}
\and {E.~Kemp}\thanksref{y}
\and {H.~de~Kerret}\thanksref{d,6}
\and {D.~Kryn}\thanksref{d}
\and {M.~Kuze}\thanksref{t}
\and {T.~Lachenmaier}\thanksref{w}
\and {C.E.~Lane}\thanksref{i}
\and {T.~Lasserre}\thanksref{k,d}
\and {C.~Lastoria}\thanksref{h}
\and {D.~Lhuillier}\thanksref{k}
\and {H.P.~Lima Jr}\thanksref{e}
\and {M.~Lindner}\thanksref{o}
\and {J.M.~L\'opez-Casta\~no}\thanksref{h,13}
\and {J.M.~LoSecco}\thanksref{p}
\and {B.~Lubsandorzhiev}\thanksref{j}
\and {J.~Maeda}\thanksref{u,m}
\and {C.~Mariani}\thanksref{z}
\and {J.~Maricic}\thanksref{i,7}
\and {J.~Martino}\thanksref{r}
\and {T.~Matsubara}\thanksref{u,8}
\and {G.~Mention}\thanksref{k}
\and {A.~Meregaglia}\thanksref{f}
\and {T.~Miletic}\thanksref{i,9}
\and {R.~Milincic}\thanksref{i,7}
\and {A.~Minotti}\thanksref{k,10}
\and {D.~Navas-Nicol\'as}\thanksref{d,h,2}
\and {P.~Novella}\thanksref{h,11}
\and {L.~Oberauer}\thanksref{v}
\and {M.~Obolensky}\thanksref{d}
\and {A.~Onillon}\thanksref{k}
\and {A.~Oralbaev}\thanksref{n}
\and {C.~Palomares}\thanksref{h}
\and {I.M.~Pepe}\thanksref{e}
\and {G.~Pronost}\thanksref{r,12}
\and {J.~Reichenbacher}\thanksref{b,13}
\and {B.~Reinhold}\thanksref{o,7}
\and {S.~Sch\"{o}nert}\thanksref{v}
\and {S.~Schoppmann}\thanksref{o,15}
\and {{L.~Scola}}\thanksref{k}
\and {R.~Sharankova}\thanksref{t}
\and {V.~Sibille}\thanksref{k,4}
\and {V.~Sinev}\thanksref{j}
\and {M.~Skorokhvatov}\thanksref{n}
\and {P.~Soldin}\thanksref{a}
\and {A.~Stahl}\thanksref{a}
\and {I.~Stancu}\thanksref{b}
\and {L.F.F.~Stokes}\thanksref{w}
\and {F.~Suekane}\thanksref{s,d}
\and {S.~Sukhotin}\thanksref{n}
\and {T.~Sumiyoshi}\thanksref{u}
\and {Y.~Sun}\thanksref{b,7}
\and {{C.~Veyssiere}}\thanksref{k}
\and {B.~Viaud}\thanksref{r}
\and {M.~Vivier}\thanksref{k}
\and {S.~Wagner}\thanksref{d,e}
\and {C.~Wiebusch}\thanksref{a,*}
\and {G.~Yang}\thanksref{c,14}
\and {F.~Yermia}\thanksref{r}
}
%
\newcommand{\affiliation}[2]{#2\label{#1}}
\institute{
\affiliation{a}{\Aachen} 
\and \affiliation{b}{\Alabama} 
\and \affiliation{c}{\Argonne} 
\and \affiliation{d}{\APC} 
\and \affiliation{e}{\CBPF} 
\and \affiliation{f}{\CENBG} 
\and \affiliation{g}{\Chicago} 
\and \affiliation{h}{\CIEMAT} 
\and \affiliation{i}{\Drexel} 
\and \affiliation{j}{\INR} 
\and \affiliation{k}{\CEA} 
\and \affiliation{l}{\Kitasato} 
\and \affiliation{m}{\Kobe} 
\and \affiliation{n}{\Kurchatov} 
\and \affiliation{o}{\MaxPlanck} 
\and \affiliation{p}{\NotreDame} 
\and \affiliation{q}{\IPHC} 
\and \affiliation{r}{\SUBATECH} 
\and \affiliation{s}{\TohokuUni} 
\and \affiliation{t}{\TokyoInst} 
\and \affiliation{u}{\TokyoMet} 
\and \affiliation{v}{\Muenchen} 
\and \affiliation{w}{\Tubingen} 
\and \affiliation{y}{\UNICAMP} 
\and \affiliation{z}{\vtech} 
\and \affiliation{aa}{\Chooz} 
}

%
%
\thankstext{1}{Now at \SUSSEX}
\thankstext{2}{Now at \ORSAY}
\thankstext{3}{Now at \Londrina}
\thankstext{4}{Now at \MIT}
\thankstext{5}{Now at \TokyoSci}
\thankstext{6}{Deceased.}
\thankstext{7}{Now at \Hawaii}
\thankstext{8}{Now at \KEK}
\thankstext{9}{Now at \Arcadia}
\thankstext{10}{Now at \LAPP}
\thankstext{11}{Now at \IFIC}
\thankstext{12}{Now at \SK}
\thankstext{13}{Now at \SD}
\thankstext{14}{Now at \StonyBrooks}
\thankstext{15}{Now at \UCB}
%


\thankstext{*}{Corresponding author, e-mail: wiebusch@physik.rwth-aachen.de}

\date{Received: date / Accepted: date}

\maketitle

\begin{abstract}

We present a search for signatures of neutrino mixing of electron anti-neutrinos with additional hypothetical sterile neutrino flavors using the Double Chooz experiment. 
The search is based on data from 5 years of operation of Double Chooz, including 2 years in the two-detector configuration.
The analysis is based on a profile likelihood, i.e.\ comparing the data 
to the model prediction of disappearance in a data-to-data comparison of the two respective detectors. 
The analysis is optimized for a model of three active and one sterile neutrino. It is sensitive in the typical mass range $\SI{5e-3}{eV^2} 
\lesssim \Delta m^2_{41} \lesssim \SI{3e-1}{eV^2} $ for mixing angles down to $\sin^2 2\theta_{14} \gtrsim \num{0.02} $. 
No significant disappearance  additionally to the  conventional disappearance related to  $\theta_{13}  $ is observed and correspondingly exclusion bounds on the sterile mixing parameter  $\theta_{14}  $ as a function  of  $ \Delta m^2_{41} $ are obtained.

\end{abstract}

\section{Introduction}

The standard model of particle physics includes three flavors of neutrinos that interact through the weak force with other  particles \cite{Tanabashi:2018oca}. The neutrino flavors are
identified by the corresponding charged lepton in charged current interactions. With the discovery \cite{Fukuda:1998mi,Ahmad:2002jz} of  neutrino oscillations \cite{Pontecorvo:1967fh,Maki:1962mu}, 
it became clear that neutrinos have mass. Currently the majority of observations is consistent with the standard picture of three mass eigenstates $(\nu_1,\nu_2, \nu_3)$  mixing with the flavor eigenstates $(\nu_e,\nu_\mu, \nu_\tau)$.
The mixing is described by  a $3\times 3 $ unitary matrix (PNMS matrix),  parametrized by three mixing angles $\theta_{12}$, $\theta_{13}$, $\theta_{23}$ as well as a CP violating phase $\delta$ and two Majorana phases if neutrinos are Majorana particles.

The neutrino experiments Double Chooz, Daya Bay, and RENO   contributed to the field by establishing the third oscillation mode that is related to the mixing angle $\theta_{13}$ \cite{Abe:2011fz,An:2012eh,Ahn:2012nd}. These experiments observe the disappearance of $\nubar_e$ from nuclear reactors by measuring the flux at different distances. The concept of multiple identical detectors has proven crucial in controlling and reducing systematic uncertainties. Today, the oscillation angle $\theta_{13}$ is the most precisely measured oscillation parameter \cite{Tanabashi:2018oca}.

There have been speculations about the existence of additional neutrinos that are non-interacting with matter, see e.g.\  \cite{Abazajian:2012ys}. 
These thoughts are supported by experimental anomalies reported by the LSND \cite{Aguilar:2001ty} and MiniBooNE \cite{Aguilar-Arevalo:2018gpe}
neutrino-beam experiments as well as the so-called reactor \cite{Mention:2011rk} and gallium 
\cite{Acero:2007su,Abdurashitov:2009tn,Kaether:2010ag}
anomalies, where the observed   $\nubar_e$  and $\nu_e$ fluxes are roughly \SIrange{5}{10}{\percent} less than the theoretical predictions. 
However, the uncertainty of those predictions remains an open question and our latest results \cite{DoubleChooz:2019qbj} indicate a possible underestimation of the reactor flux prediction.
Though this deficit is marginally compatible with the uncertainty of the flux prediction, it could be also interpreted as disappearance due to oscillation with additional neutrino states.
Recently, the Neutrino-4 collaboration has reported \cite{Serebrov:2018vdw} indications of a spectral distortion at short baseline to the reactor that would be consistent with the oscillation hypothesis. This result is subject of ongoing
discussions \cite{Danilov:2018dme,Almazan:2020drb,Serebrov:2020yvp}. Particularly it
 has been reviewed in \cite{Coloma:2020ajw} considering the validity of the Wilks’ theorem, thus resulting in a reduced significance. Note that in this paper we report a very similar effect of reduced significance with respect to Wilks' theorem in our measurement.
From a phenomenological perspective it is important to emphasize that consistency of all today’s global data within a single simple solution remains an unsettled open debate, see e.g.\  \cite{Dentler:2018sju}.

The simplest extension of the standard oscillation picture is a $3+1$ model \cite{Abazajian:2012ys}.
Though this model cannot consistently explain all experimental anomalies, its few parameters make it well suited as a benchmark model in  the following discussions.
Here, one additional sterile, i.e.\ not weakly interacting, neutrino mixes with  the three active neutrino states. This results in an additional mass state $m_4$ and an  extension of the mixing matrix to $4\times 4$ with the additional parameters $\theta_{14}$, $\theta_{24}$, $\theta_{34}$, and  additional CP violating phases.

\begin{figure}[htbp]
    \centering
    \includegraphics[width=\columnwidth]{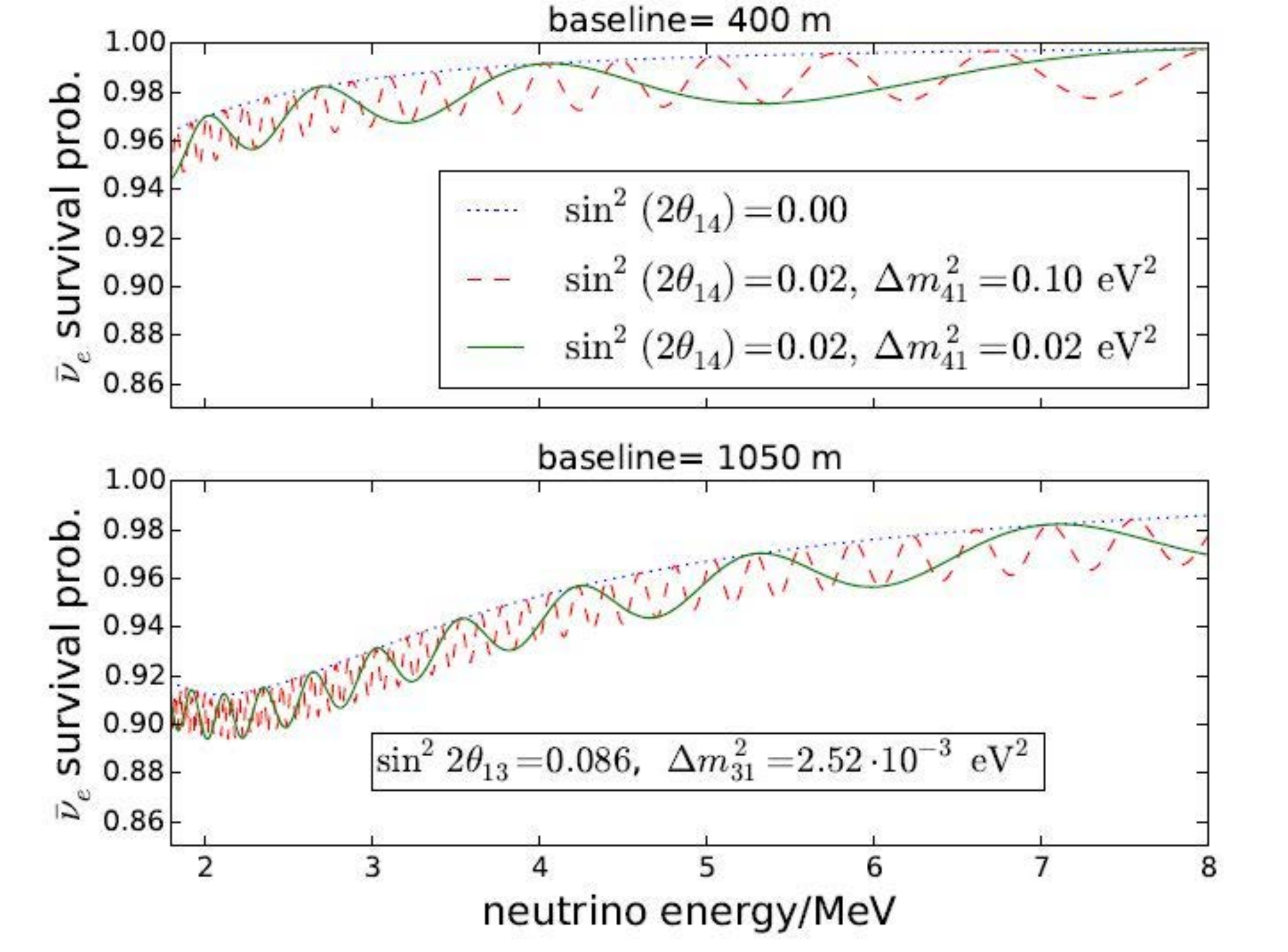}
    \caption{Survival probability of reactor $ \nubar_e $ as a function of the energy for the baselines of the ND (top) and FD (bottom) for different benchmark oscillation parameters $\theta_{14}$ and $\Delta m^2_{41}$. 
    The dotted line corresponds to the no-sterile case, where the survival probability is governed by the conventional $\theta_{13} $ oscillation. The dashed and solid lines show two different examples of sterile mixing.
    }
    \label{fig:experimental_signature}
\end{figure}

In this picture, a non-zero mixing of reactor $\nubar_e$ with a sterile neutrino will result in a disappearance, superimposed to the standard oscillation related to $\theta_{13}$. 
Assuming small mixing and 
baselines relevant for the Double Chooz experiment, only the parameters $\theta_{14}$ and the difference of squared masses  $\Delta m^2_{41} \equiv m_4^2 - m_1^2$ are relevant \cite{Kopp:2013vaa}, and the survival probability of $\nubar_e$ as a function of distance $L$ and energy $E$  can be approximated by 
\begin{multline}
       P_{\nubar_e \rightarrow \nubar_e}\left(E, L\right) 
    \:\approx  1   - \sin^{2}\left(2\theta_{13} \right)\sin^{2}\left(\SI{1.267}{MeV \over  eV^2 \meter}  \cdot \frac{\Delta m^{2}_{ee} L}{E}\right)  \\
     \: - \sin^{2}\left(2\theta_{14} \right)\sin^{2}\left(\SI{1.267}{MeV \over eV^2 \meter}  \cdot\frac{\Delta m^{2}_{41} L}{E}\right) 
     \label{eq:oscillation}
\end{multline}
The first sine term 
corresponds to the disappearance  related to the standard $\theta_{13}$ mixing while the
second sine term 
describes the additional disappearance due to the mixing with the sterile neutrino state. The term $ \Delta m^{2}_{ee} $ is a shorthand for $ \cos^2 \theta_{12} \, \Delta m^2_{31}  + \sin^2 \theta_{12} \, \Delta m^2_{32} $. 

The effect is displayed in Fig.~\ref{fig:experimental_signature} for  baselines of \SIlist{400;1050}{m} corresponding to the average distances of the nuclear reactors
to the two Double Chooz detectors.
    The existence of sterile neutrinos with non-zero mixing leads to the additional disappearance superimposed on the conventional oscillation. The amplitude of this oscillation is given by the parameter
    $ \sin^{2}\left(2\theta_{14} \right) $.
    The oscillation frequency seen in the energy-dependence is proportional to the difference of squared masses.
    For mass differences of $\Delta m^2_{41} \gg \SI{0.1}{eV^2} $, oscillations become fast.  Given the experimental energy resolution, they become eventually  indistinguishable from a glo\-bal normalization change. 
    Similarly, 
    for small mass-square differences  $\Delta m^2_{41} \approx \Delta m^2_{ee} 
        \simeq \SI{2.5e-3}{eV^2} $
     the disappearance becomes indistinguishable
     from the conventional oscillation with $\theta_{13}$.
Note, that the above approximation is only used for illustrative purposes and for all numerical calculations in this analysis we use the full four-flavor propagation code \texttt{nuCraft} \cite{Wallraff:2014qka}.

The position of the two Double Chooz detectors has been optimized for the measurement of $\theta_{13}$ assuming 
$\Delta m^2 \approx \SI{2.5e-3}{eV^2}$. 
For an energy range of detected reactor neutrinos
 between about from \SIrange{1}{8}{MeV} and the two baselines of \SIlist{400;1050}{m}, the probed $L/E$ range for the disappearance of $\nubar_e $ is approximately \SIrange{50}{1000}{m \per MeV}.
For larger mass differences, shorter baselines are desirable in order to observe the un-oscillated flux with a near detector. This is realized by short-baseline experiments, Bugey-3 \cite{Declais:1994su} and more recently  DANSS \cite{Alekseev:2018efk},
NEOS \cite{Ko:2016owz}, 
Neutrino-4 \cite{Serebrov:2018vdw}, PROSPECT \cite{Ashenfelter:2018iov}, 
SoLID \cite{Abreu:2017sit}, 
and STEREO \cite{Almazan:2018wln,AlmazanMolina:2019qul}, 
that target 
mass-square differences on the \si{eV^2} scale.
The probed $L/E$ range for these experiments is typically \SIrange{1}{20}{m \per MeV}.
Therefore the here presented search is complementary in probed $L/E$ as well as lower 
probed mass-square differences below \SI{0.1}{eV^2}; see also \cite{Esmaili:2013yea}.

\section{Experimental setup}

\begin{figure}[htbp]
    \centering
    \includegraphics[width=\columnwidth]{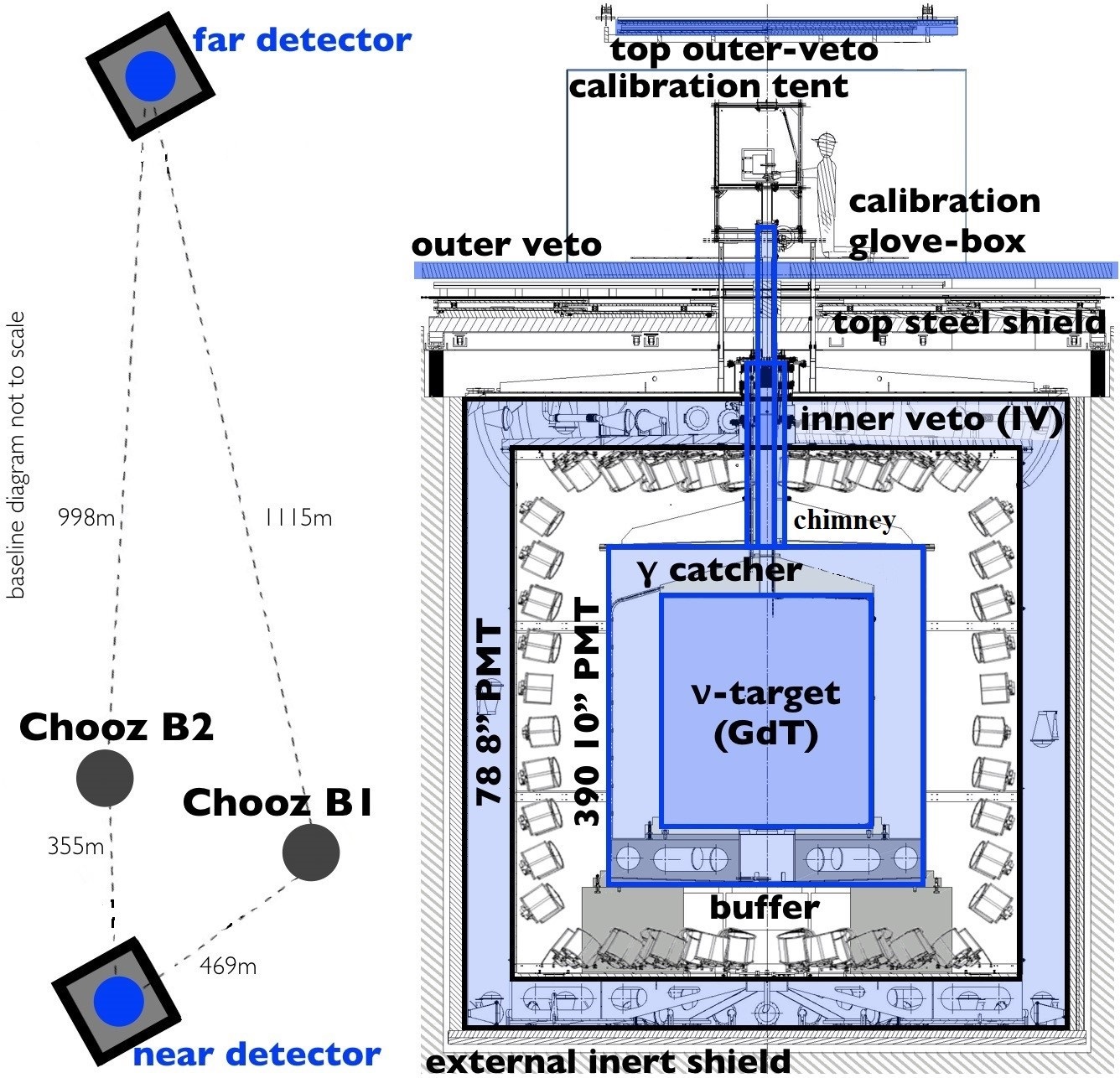}
    \caption{The Double Chooz experiment. Left: arrangement of the two detectors far and near with respect to the nuclear reactors. Right: Design of a Double Chooz detector. Figure modified from \cite{DoubleChooz:2019qbj}.  }
    \label{fig:experiment}
\end{figure}

The Double Chooz experiment  consists of two nearly identical  gadolinium-doped liquid scintillator detectors \cite{Ardellier:2006mn} located close to the Chooz-B nuclear power plant, see Fig.~\ref{fig:experiment}. 
The power plant consists of two nuclear reactors of type N4,  \SI{165}{m} apart with a thermal power of about \SI{4.25}{\giga \watt} each. The far (near) detector
is located underground with an overburden of about \SI{300}{m} (\SI{120}{m}) water equivalent  at  a distance of \SI{1115}{m} and \SI{998}{m} (\SI{469}{m} and \SI{355}{m}) 
to the reactor cores. 

Details of the detectors are described in \cite{Abe:2011fz,Abe:2012tg,Abe:2014bwa,DoubleChooz:2019qbj}.
The detectors are constructed in an onion-like structure with a  central detector made of four concentric
cylindrical tanks. 
The innermost acrylic vessel contains \SI{10.3}{m^3} gadolinium loaded liquid scintillator called the $\nu $-target. The $\nu $-target is surrounded by the $\gamma$-catcher, filled with \SI{22.5}{m^3} liquid scintillator without gadolinium loading. Both central volumes serve as the neutrino target. A neutrino interacting
in the target by  inverse beta decay ($\nubar_e+ p \to e^++n $) \cite{Vogel:1999zy} produces  the characteristic signature of a delayed coincidence well known since the early days of neutrino experiments \cite{Cowan:1992xc}. 
This is formed by  a prompt signal from the positron and its annihilation 
and then the delayed signal from the capture of the thermalized neutron by either gadolinium or hydrogen. Though increasing the rate of accidental background events, the use of both types of captures in a combined data set greatly enlarges the sensitive volume and thus the statistics of detected neutrinos as well as reducing some of the systematic uncertainties. This technique, called total neutron capture, has been developed by the Double Chooz experiment for the most recent $\theta_{13} $ analysis 
\cite{DoubleChooz:2019qbj} and is applied also for this analysis. 

The central target volumes are  surrounded by a buffer volume filled with mineral oil, shielding the inner volume from radioactivity, partly from  \num{390} 10-inch PMTs that are installed on the inner wall of the stainless steel buffer tank and  observe the target. Optically separated from these inner volumes is the inner veto. That is a \SI{50}{cm} thick cylindrical volume filled with liquid scintillator and  equipped with \num{78} 8-inch PMTs.
It actively shields the inner detector 
by tagging cosmic-ray induced muons, gammas, and neutrons from outside the detector.
Shields of \SI{15}{cm} thick demagnetized steel (\SI{1}{m} water) surround the inner veto of the far (near) detector, suppressing external gamma rays.    
A chimney in the top center allows deploying radioactive sources for calibration. Above the detector is the outer veto detector that adds to the shielding and allows for evaluating the efficiency of the inner veto detector.

Several key aspects of the Double Chooz experiment are important to this analysis. A main goal is avoiding dependencies on absolute predictions of the neutrino flux from the reactors as well as detection efficiencies. Therefore we perform
a direct comparison of the event rates measured in the two identical 
detectors, in the following referred to as \emph{data-to-data} approach.
This results in the cancellation of most reactor flux related uncertainties as well as detection efficiencies and some of the background uncertainties in the measurement of $\nubar_e$ disappearance. 
Furthermore, due to the presence of only two, relatively close reactor cores, the geometry
constitutes well defined baselines from the reactors to the detectors
which is important for testing faster oscillation modes than the $\theta_{13} $ oscillation (see Fig. \ref{fig:experimental_signature}).
 The two detectors  are situated close to the so-called iso-flux line,
where the ratio of neutrino fluxes from the two reactors is the same for both detectors, i.e. the relative contribution from the two reactors is very similar in the two detectors, further reducing  the reactor 
uncertainty. 
Another important aspect is that we include measurements when one of the reactors or even both reactors were switched off. These data allow for directly measuring the backgrounds and their spectral properties
\cite{Abe:2012ar,deKerret:2018fqd}. In this analysis, the  data from these off-reactor phases are used to construct templates of the energy distribution of backgrounds as well as the uncertainties of these templates for the fit to data.
 Additionally, the total rate is used to constrain the background rates.

Experimental backgrounds include   uncorrelated backgrounds,
where a single event appears in a random coincidence with another  event,
as well as correlated backgrounds that mimic both the prompt and the delayed event.
The dominant sources of uncorrelated backgrounds are 
natural radioactivity and instrumental noise such as spontaneous light emission in the PMT bases of the far detector \cite{Abe:2016ixi}. Correlated backgrounds are mostly caused by
secondary products from cosmic ray air induced atmospheric muons
that pass close or through the detectors.
Muons reaching the detector
are detected with high efficiency and cause an active veto of \SI{1.25}{\milli \second} duration. However, background events
arise by (i) fast neutrons from interactions in the rock close to the detector entering the neutrino target, (ii) long lived isotopes, in particular \ce{^9Li} 
\cite{deKerret:2018fqd}, that undergo $\beta $-decays followed by neutron emission, and (iii) low energy stopping muons that enter the detector through the chimney and decay by emission of a Michel electron. All these backgrounds are considerably reduced during the data selection and the remainder are measured with specific methods and in dedicated campaigns, e.g.\ during reactor-off phases.

The data  of this analysis are identical to the selection described in \cite{DoubleChooz:2019qbj} and are separated into three data sets. The first (FD-I) has been collected with the far detector prior to commissioning of the near detector and consists of \num{455.21} days of dead and down-time corrected livetime, collected between April 2011 and January 2013. The second set (FD-II) has been collected with the far detector during operation of both detectors and consists of \num{362.97} days of livetime collected between January 2015 and April 2016.
The third set (ND) are the data collected during the same period with the near detector and corresponds to \num{257.96} days of livetime. Note that the effective livetime of the ND data is reduced with respect to the FD-II data, because the larger muon rate in the near detector causes a larger dead-time due to vetoing.
While the previously described  data has been collected during operation of at least one reactor, additionally \num{7.16} days of livetime with both reactors switched off 
during the FD-I phase
are used to determine the total rate of background events.

\section{Analysis Method}

The analysis is based on a profile likelihood ratio (see e.g.\ G.\ Cowan in \cite{Tanabashi:2018oca}) that has already been  exploited by Double Chooz for a measurement of $\theta_{13} $ in \cite{StSchopp_PHD} and has also been used internally to confirm the result in \cite{DoubleChooz:2019qbj}. 
The test statistic is defined as the ratio of  maximum likelihoods for tested model parameters
  $\vec{\eta} =\{ \sin^2 2 \theta_{14} , \Delta m^2_{41} \}$ with respect to the globally largest likelihood value which is found for the parameters
  $\hat{\vec{\eta}} =\{ \hat{\sin^2 2 \theta_{14}} , \hat{\Delta m^2_{41} }\}$.
  This defines the test statistic for the given data set  $\vec{x}$ and model parameters  $\vec{\eta}$
\begin{equation}
    \lambda(\vec{x},\vec{\eta}) = - 2 \cdot \ln 
    \frac{\sup \mathcal{L} (\vec{x} | \vec{\eta} ,\vec{\xi}) }{\sup \mathcal{L} (\vec{x} | \hat{\vec{\eta}},\hat{\vec{\xi}} )  }
    = -2 \Delta \ln (\mathcal{L})
\end{equation}
In addition to the two model parameters
 $\vec{\eta} $ that describe a sterile neutrino signal, the reactor fluxes, detector responses, systematic uncertainties and backgrounds are modeled by a total number of \num{298} additional and partly correlated  parameters $\vec{\xi} $ (see below for details).  
These parameters are treated as nuisance parameters in the fit. They are optimized separately for each respective signal hypothesis with  
${\vec{\xi}} $ representing those nuisance parameters that maximize the local likelihood for the tested $\vec{\eta} $.

For the test of a potential oscillation signal from sterile neutrinos, we compare
the best-fit standard 3-flavor model (null hypothesis, $\vec{\eta}_0$),
described by the two parameters  $\sin^2 2 \theta_{14} =0 $ and $ \Delta m^2_{41} =0$,
to the globally best fit 
 3+1 sterile neutrino model 
(signal hypothesis) for the parameters
 $\hat{\vec{\eta}} $
 that maximize the likelihood of the data $\vec{x}$. Note that specifically the null hypothesis $\vec{\eta}_0$ is degenerate with respect to the two parameters $\vec{\eta}$ because only one of them fixed to zero is sufficient to model a no-oscillation signal. Furthermore, $\vec{\eta}_0$ is a special case, nested within the parameter space of the signal hypothesis  resulting in $\lambda(\vec{x} ,\vec{\eta}_0)\ge 0$.


The likelihood itself is implemented as a product of multiplicative terms with the Poissonian likelihoods $\mathcal{P}(n_i,\mu_i)$ of the observed number of events $n_i$ in the energy bin $i$ in all three data sets $d \in \{ ND , FD-I , FD-II \}$ 
multiplied with Gaussian prior functions $\mathcal{G} $  on external nuisance parameters
\begin{equation} 
\begin{split}
\mathcal{L} (\vec{x} | \vec{\eta} ,\vec{\xi}) 
&=
\prod_{d\in\{ND,FD-I,FD-II\}} \ 
\prod_{i\in [E_{min}\dots E_{max} ]}
\mathcal{P}(n_{d,i},\mu_{d,i}(\vec{\eta},\vec{\xi})) \\
& \cdot \mathcal{P}(n_{off},\mu_{off}(\vec{\xi})  ) \\
 &\cdot \prod_{a\in \vec{\xi} }
\left ( \mathcal{G} (a, a_0, \sigma_a) \right ) \\
 &\cdot \prod_{\vec{b}\in \vec{\xi} }
\left ( \mathcal{G} ( (\vec{b}-\vec{b}_0)^T \mathbf{V}_b^{-1} (\vec{b}-\vec{b}_0) ) \right )
\end{split}
\end{equation}
Here $\mu_{d,i} (\vec{\eta},\vec{\xi})$ denotes the summed bin expectations of signal and backgrounds  as a function of the model parameters.  The second  term is the Poisson probability of the observed event number during the reactor-off phases for the background
expectation as a function of the nuisance parameters. The third term describes Gaussian priors for all  single, uncorrelated nuisance parameters $a$ with the expectation $a_0$ and the uncertainty $\sigma_a$. The fourth term describes Gaussian priors for all  nuisance parameters $\vec{b} $ that are correlated, described by the expectation $\vec{b}_0 $  and the covariance matrix $\mathbf{V}_b $.

The data are binned for each of the three sets in 38 bins between \SIrange{1}{20}{MeV} with custom bin sizes.
The region up to  \SI{8}{MeV} which is dominated by measured reactor $\nubar_e $ has 28 bins of \SI{0.25}{MeV} size.
Above \SI{8}{MeV}, bins are background dominated but are included in the fit as they allow for  constraining the 
background rates. Due to the lower statistics, larger bin sizes are used. These
are \num{4} bins of \SI{0.5}{MeV} size between
\SIrange{8}{10}{MeV}, where rare isotopes (\ce{^9Li}) dominate and \num{4} bins of \SI{2}{MeV} size between
\SIrange{12}{20}{MeV}, where fast neutrons  dominate. In the intermediate region 
\SIrange{10}{12}{MeV},  \num{2} bins of  
\SI{1}{MeV} size are used.

\begin{figure}[htbp]
    \centering

\includegraphics[width=.33\textheight]{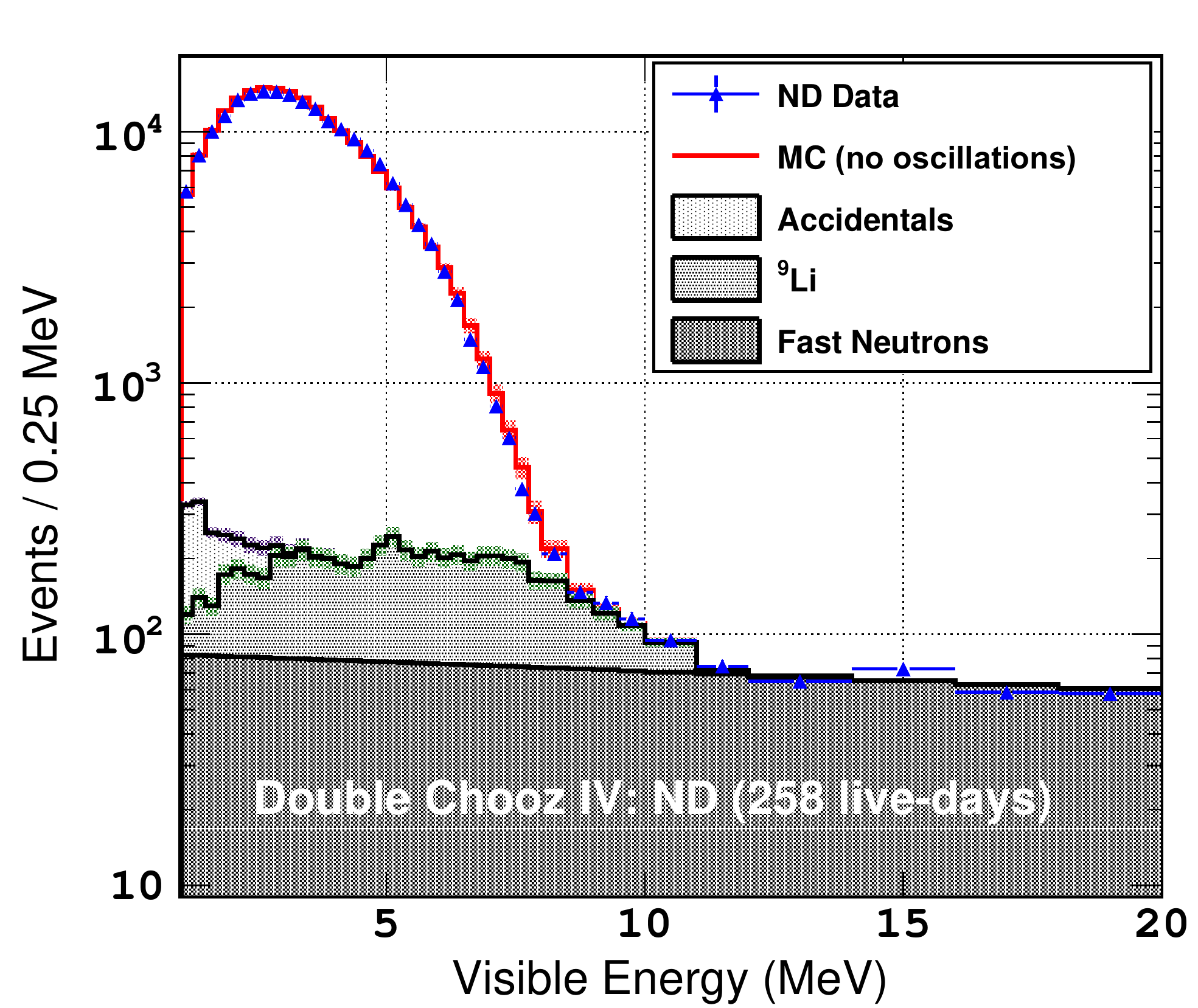}
\includegraphics[width=.33\textheight]{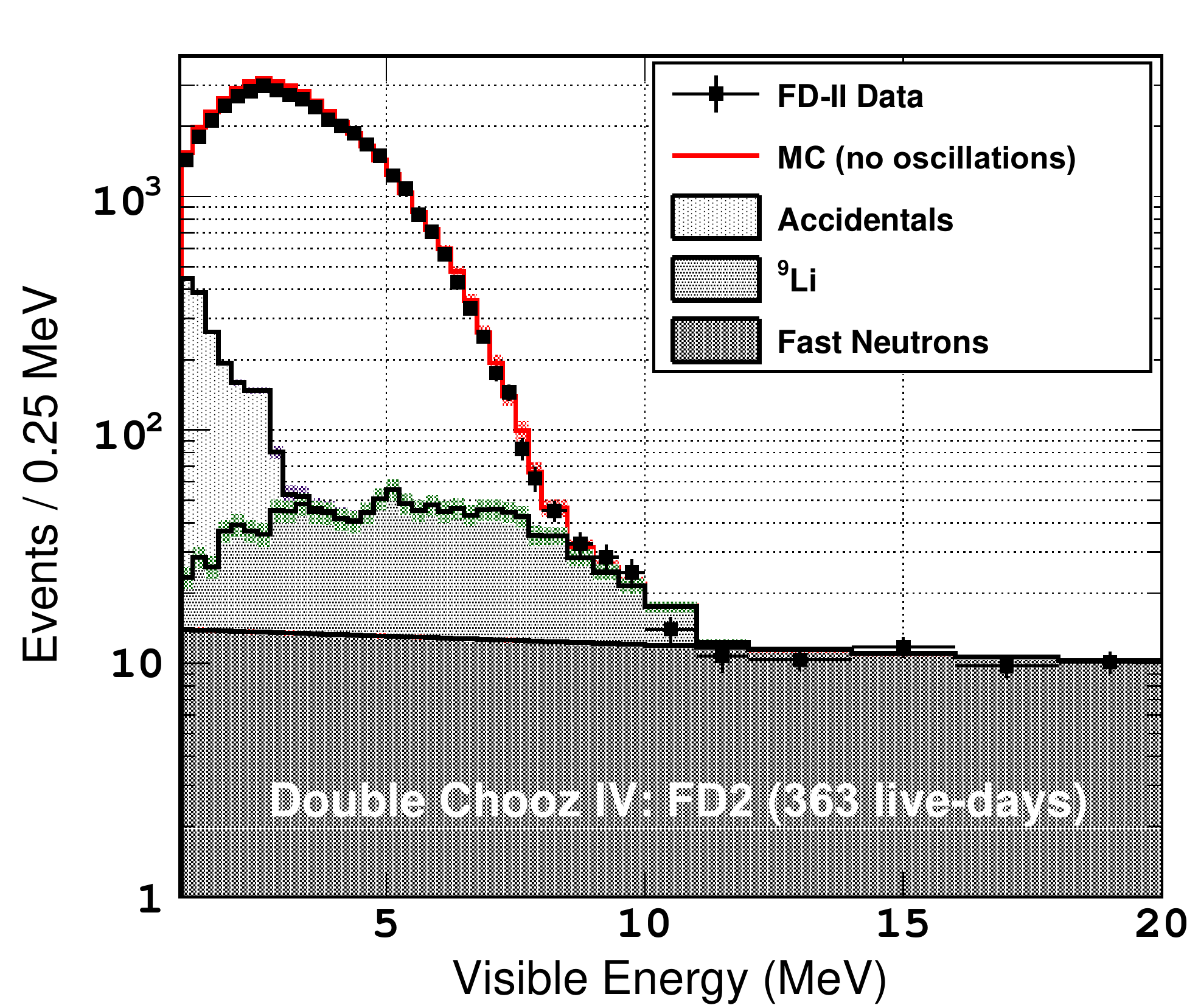}
\includegraphics[width=.33\textheight]{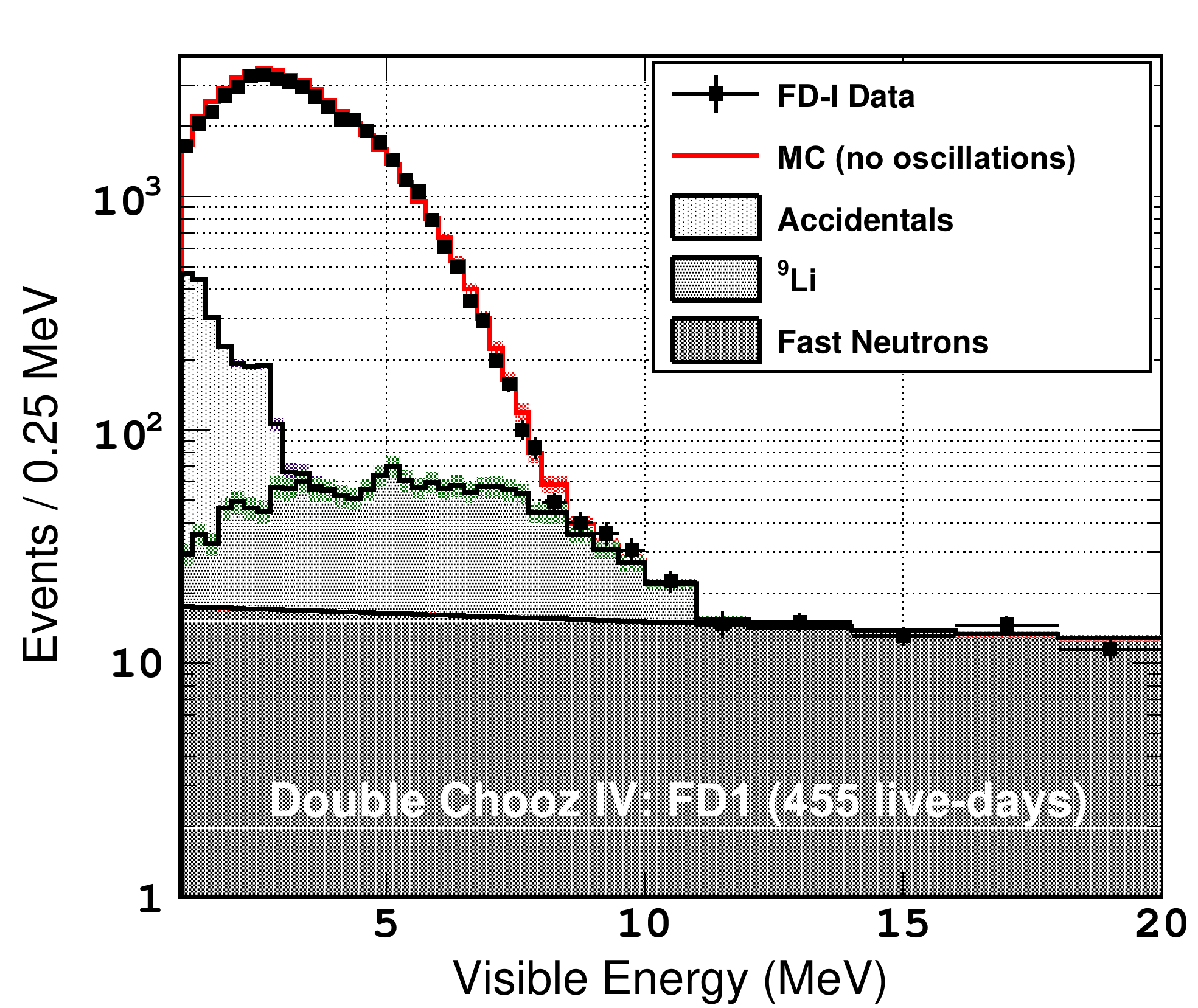}

    \caption{Visible energy distributions of the prompt events in the final data set.  The ND (top) data is plotted with blue triangles  and the FD-II (middle) and FD-I data are displayed as black
squares. The different background model contributions are shown as stacked histograms where green indicates
the long-lived isotopes (lithium) background, blue the accidental background and gray the fast neutron and stopping muon background. The red line indicates the total prediction from reactor models assuming no oscillations including the backgrounds.
    }
    \label{fig:experimental_data}
\end{figure}

Systematic uncertainties are modeled by the following nuisance parameters $ \vec{\xi}$ in the analysis (more details are given in \cite{DeniseHellwig_PHD}):
\begin{itemize}
\item 
The normalizations of the reactor flux expectation for each energy bin are free fit parameters. This approach is independent of  existing reactor flux predictions and the normalizations are only constrained by the data-to-data comparison of rate and shape of the data in each detector. This way, known discrepancies of  reactor flux models \cite{DoubleChooz:2019qbj,An:2016srz,Kwon:2017nzf,Ko:2016owz}, being independent of the baseline,  do not bias the fit, however, at the price of a slightly reduced sensitivity. 
The basis of the above approach is a large correlation in the observed reactor flux for the three data sets FD-I, FD-II, ND. 
Because of different running times, this assumption is only approximate, (\SI{99.75}{\percent} for FD-II and ND,
\SI{93.20}{\percent} for FD-I and FD2,
\SI{93.10}{\percent} for FD-I and ND).
Therefore, we model additional constraints on the normalization of each energy bin of the three data sets with a total of $3\times 41$ reactor flux parameters between \SIrange{1}{11.25}{MeV}. The number of  parameters is determined by the greatest common divisor of the bin widths to create a uniform binning. These bins form the basis of an area conserving spline, which is energy corrected and later integrated over in the original binning. These parameters are correlated between the data sets with the above correlation factors and additionally we allow for uncorrelated shape deviations with a $41\times 41$ covariance matrix for each data set, that is determined from the reactor flux prediction. 
\item The conventional oscillation parameters $ \sin^2 2\theta_{13} $ and $ \Delta m^2_{ee} $ are free parameters.
While $ \Delta m^2_{ee} $  is seeded with the global best value from \cite{Patrignani:2016xqp} and is constrained with a prior  corresponding to its uncertainty,
$ \sin^2 2\theta_{13} $  is left unconstrained. The latter ensures that  assumptions about the
value, which has itself been largely determined  in reactor neutrino experiments, cannot introduce a bias. By this, $ \sin^2 2\theta_{13} $ can acquire a different best-fit value for $\theta_{14} \ne 0$.

\item Backgrounds are modeled with free parameters for rate and shape.
The shape of the contribution from rare isotopes (\ce{^9Li}) is assumed identical between the three data sets.
It has been determined experimentally by a dedicated selection of events correlated in time and space with tagged muon events \cite{Abe:2014bwa} and it is modeled with \num{38} shape parameters. The rate is assumed identical for FD-I and FD-II but is different for ND. Both total rates are not constrained by a prior but are determined by the data  as free parameters during the fit.

The rates and shapes of accidental backgrounds have been determined by time-scrambled experimental data for each data set and are individually modeled by \num{38} parameters for the shape and one parameter for the rate. These parameters are  assumed uncorrelated for the three data sets, to account for changes in data taking over time and differences in the detectors but are constrained with a prior that reflects the uncertainty in the determination of the rates.

The fast neutron and stopping muon backgrounds are modeled as $ R(E) = R_0 (a\cdot E + b\cdot \exp(-\lambda \cdot E) $. Here, $R_0$ is the total rate and the three shape parameters $\lambda$, $a$, and $b$ are further constrained by the normalization. The shapes are assumed to be fully correlated between the data sets, while the rates   are the same for for FD-I and FD-II but independent  for ND.

A special case is the small constant rate  of $\nubar_e $ from the reactor fuel that has been determined during FD-I reactor-off phases to \SI{0.58+-0.17}{\per \day}. As these neutrinos undergo the same oscillation, this is modeled in the fit with the nominal oscillated shape expectation for $\nubar_e $ and the rate is constrained by a prior corresponding to this reactor-off rate.


\item The uncertainties in the detector response are modelled identically to \cite{DoubleChooz:2019qbj} by second order polynomials. 
They take into account the non-linearity of the visible energy response of the scintillator, the non-uniformity within the detector, and the charge non-linearity of the photomultiplier and electronics response. 
After analyzing the correlations of these effects
where we assume the  energy response of the scintillator to be fully correlated but the 
other effects to be uncorrelated between the data-sets, the \num{9} polynomial coefficients can be expressed by  \num{7} independent parameters.
In addition to the energy response, the total detection efficiency is subject to uncertainty, dominated by the uncertainty of the total target mass. This is modeled by a total of three constrained and partly correlated parameters.

\end{itemize}
The resulting expectations of reactor $\nubar_e $ as well as the backgrounds for the default model are shown in Fig.~\ref{fig:experimental_data}  in comparison
to the experimental data for all three data sets. In addition to the above parameters we have tested additional uncertainties but their effect was found to be negligible. In particular it was shown that the choice of mass ordering has no relevant impact on the 
analysis.

The above fit has been extensively tested.
These tests include a detailed validation of the $\theta_{13} $ fit in the absence of a sterile signal that was found in good agreement to the published standard analyses of Double Chooz. Here, the relative impact of each systematic uncertainty has been evaluated by performing fits excluding the corresponding  nuisance parameter (N-1) or fits including exclusively this parameter on top of statistical uncertainties (stat+1). All resulting uncertainties have been found in 
good agreement with the standard analysis \cite{DoubleChooz:2019qbj}.

For the validation of the detection of a sterile signal, 
studies of pseudo experiments with injected signal and 
blind data-challenges have been performed. Furthermore, the impact of each systematic parameter and other experimental effects, such as the spectral distortion at \SI{5}{MeV} have been tested.
Here, it was verified that the fit results in an unbiased estimation of the parameters
$\sin^2 2\theta_{14} $ and  $\Delta m^2_{41}$.

\section{Test Statistic}

\begin{figure}[htbp]
 \centering
    \includegraphics[width=\columnwidth]{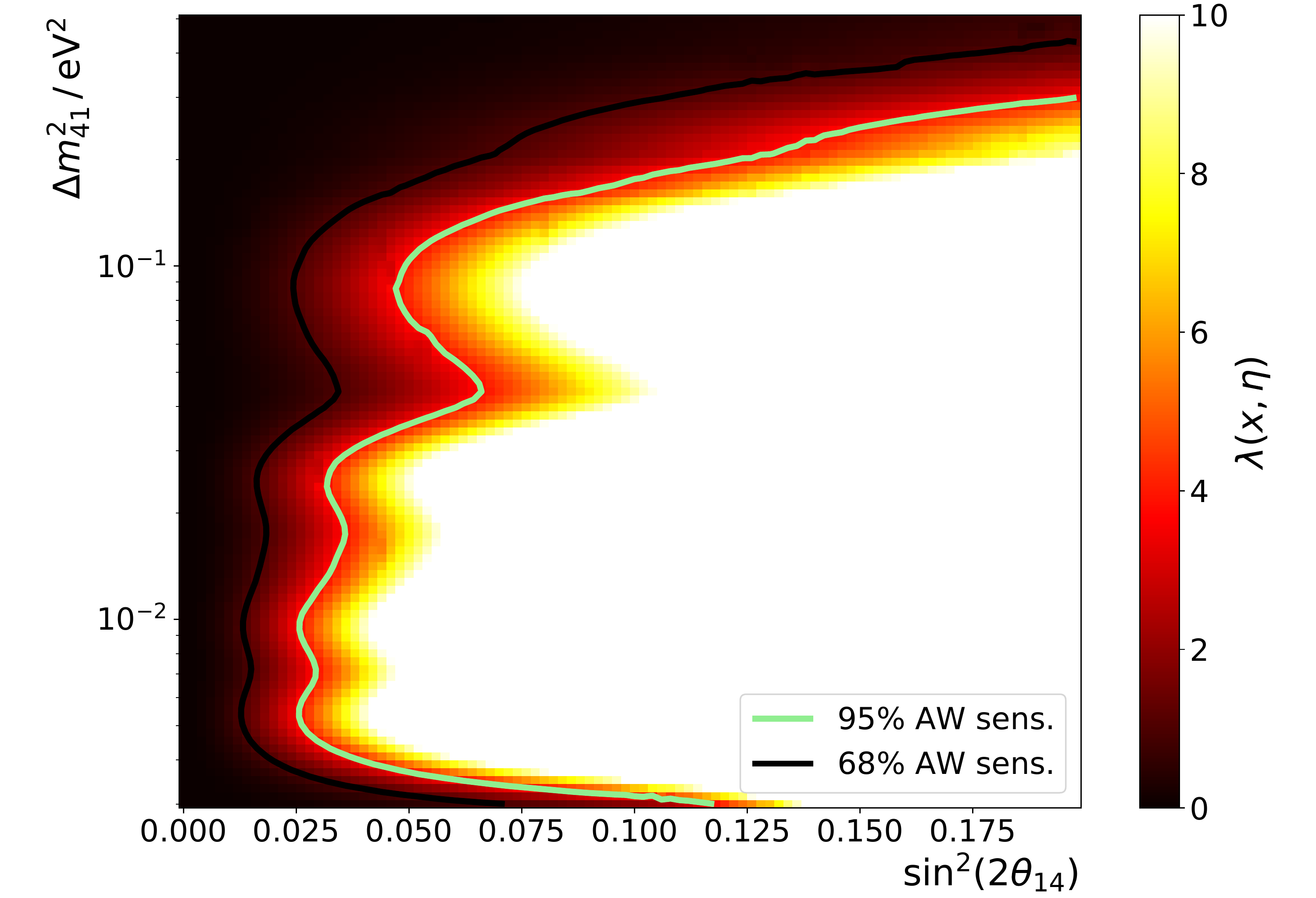}
    \caption{Test statistic $\lambda (\vec{x},\vec{\eta}) $ for an Asimov data set $\vec{x} $ of the null hypothesis for a scan of the signal parameter space. 
     All values of  $\sin^2 2\theta_{14} {=} 0$ represent the null hypothesis of no-sterile oscillations and correspondingly  $\lambda =0 $ for the Asimov data set.
    The color scale is clipped at  $\lambda =10 $.
     The lines represent the \SI{68}{\percent} and 95\% sensitivity (see text) for constraining  $\sin^2 2\theta_{14} $ as a function of $\Delta m^2_{41}$.}
    \label{fig:teststatistics}
\end{figure}

The maximum likelihood is numerically obtained by minimizing the negative $\log (\mathcal{L})$. However, finding  the global minimum and $\hat{\vec{\eta}}$ is numerically challenging because
the fit does not converge for arbitrary combinations of initial signal and nuisance parameters to the global minimum.
Therefore, 
the full phase space of signal parameters $\vec{\eta}$ is scanned by performing a numerical fit of the  parameters $\vec{\xi}$ for each scan point.
The result of such a scan 
 is shown in Fig.~\ref{fig:teststatistics}
for an Asimov data set \cite{Cowan:2010js} based on  Monte Carlo simulations of the null hypothesis of only standard oscillations. As the Asimov data set represents the mean expectation for this hypothesis, we thus find $ \lambda (\vec{x}) =0 $ for  $\sin^2 (2\theta_{14} ) =0$ corresponding to the injected null hypothesis.

As noted above, the null hypothesis 
is  a special case nested within the more general signal hypothesis. 
The test statistic thus allows for a hypothesis test for a sterile signal i.e. non-zero $\vec{\eta}$ with respect to the no-sterile case  $\vec{\eta}_0 =\vec{0}$ based on the likelihood ratio.
If applied, Wilks' theorem \cite{Wilks:1938dza} would predict that  the test statistic $TS =\lambda (\vec{x},  \vec{\eta}_0 )$ follows a $\chi^2$ distribution with two degrees of freedom corresponding to the difference in 
degrees of freedom of the signal  and null hypotheses.
However, the preconditions for  Wilks' theorem
are not fulfilled.
First, the two parameters $\sin^2 2\theta_{14}  $ and $\Delta m^2_{41}$
are degenerate in case of the null hypothesis. Any combination of these 
with one of the two parameters equal  to zero
is sufficient for fulfilling the null hypothesis  even if the other parameter has a non-zero value. In many practical applications one 
can accommodate the problem by introducing an effective degree of freedom $1 \le n_{eff}  \le 2 $ and the value of $n_{eff}$   can be estimated by pseudo experiments
with the method introduced by Feldman and Cousins \cite{Feldman:1997qc}.
Secondly, 
the expectation value of 
partial derivatives 
with respect to the parameters 
$||\langle \frac{\partial^2 \mathcal{L} (\vec{x} | \vec{\eta})}{\partial \eta_i \ \partial \eta_j} \rangle || $ 
should form a positively definite matrix.
Due to the oscillatory structure of the signal hypothesis, this is not the case here.
A data fluctuation in any of the energy bins can be better described by some signal hypotheses that correspond to such an oscillatory pattern in the detectors.
As a matter of fact, multiple, very different
signal parameters can lead --- within the experimental resolutions --- to similar patterns. 
In an illustrative picture, for a statistical fluctuation of the experimental data bins in energy, multiple different combinations of signal parameters allow for a slightly improved description of the data with respect to  the null hypothesis.
As a result, multiple
minima of the test statistic can be found within the signal parameter space.
However, the existence of several minima implies that the above matrix of derivatives is zero in some points of the parameter space.

\begin{figure}[htbp]
    \centering
    \includegraphics[width=\columnwidth]{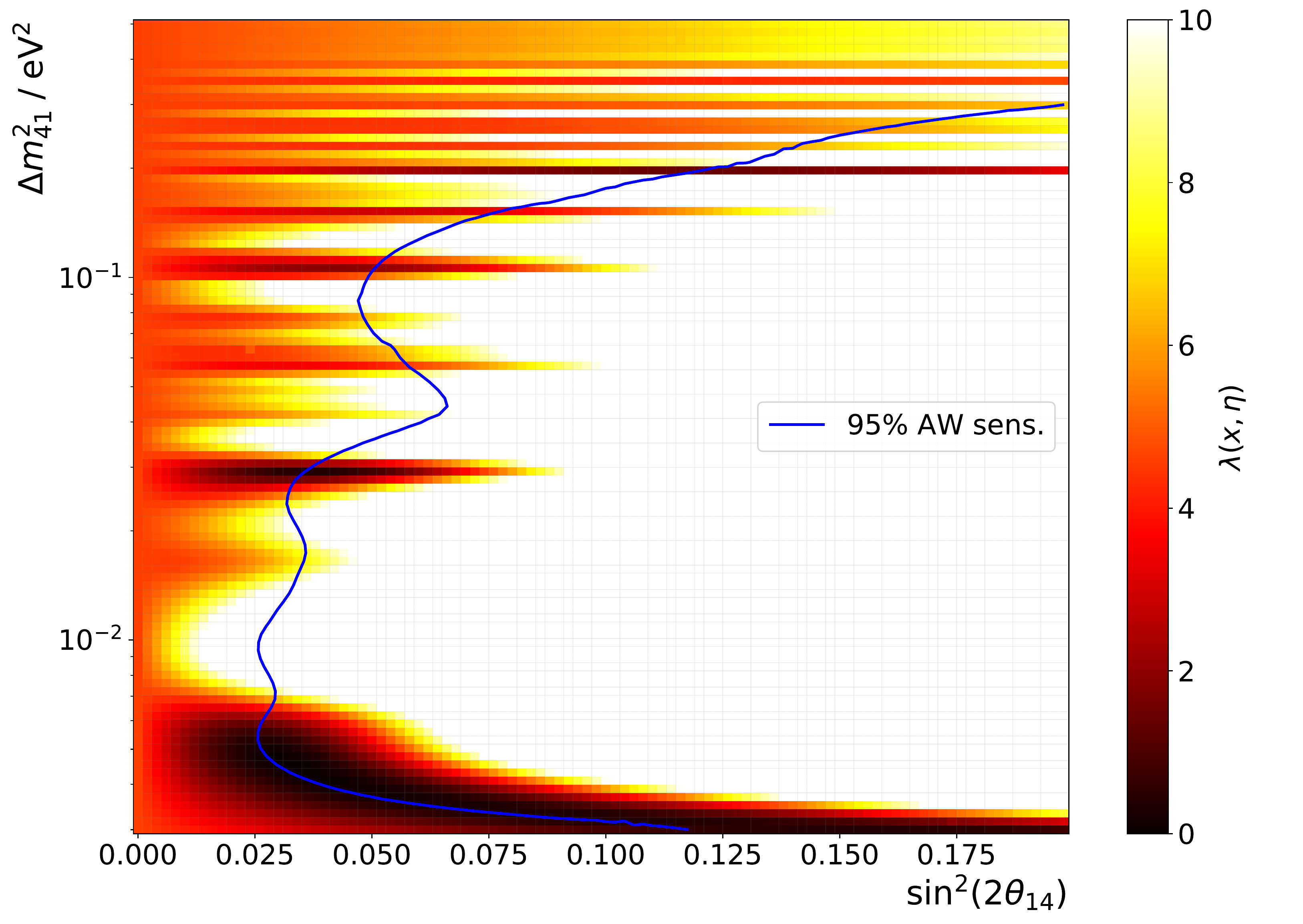}
    \caption{Example analysis of a pseudo data set representing the null hypothesis. The data set was generated with Poissonian fluctuations from a Monte Carlo data set. The blue line represent the 95\% sensitivity (as defined in the text) for constraining  $\sin^2 2\theta_{14} $ as a function of $\Delta m^2_{41}$.
    }
    \label{fig:toyexperiment}
\end{figure}

As verification of the above discussion, Fig.~\ref{fig:toyexperiment} shows an example analysis for a  pseudo data set that was generated from a Monte Carlo simulation of the null hypothesis.
The occurrence of multiple minima of the test statistic is well visible. As apparent features, these minima are horizontally elongated and thus correspond to a fixed 
value of $\Delta m^2_{41}$. Repeated pseudo experiments show similar features with, however,  different number of minima and locations in each experiment. 
This supports the interpretation that for each possible  statistical representation of the null hypothesis, multiple signal hypotheses can be found that describe the observed data slightly better than the average expectations from the null hypothesis. Each such solution requires a  fixed oscillation length and is usually found close to the sensitivity-line  beyond the region where a stronger signal would likely cause a more significant observation.
More details on these observations can be found in \cite{DeniseHellwig_PHD}. 
We note that this 
has been independently discussed in \cite{Agostini:2019jup} for short-baseline sterile neutrino searches and very recently in \cite{Coloma:2020ajw}.

\begin{figure}[htbp]
    \includegraphics[width=\columnwidth]{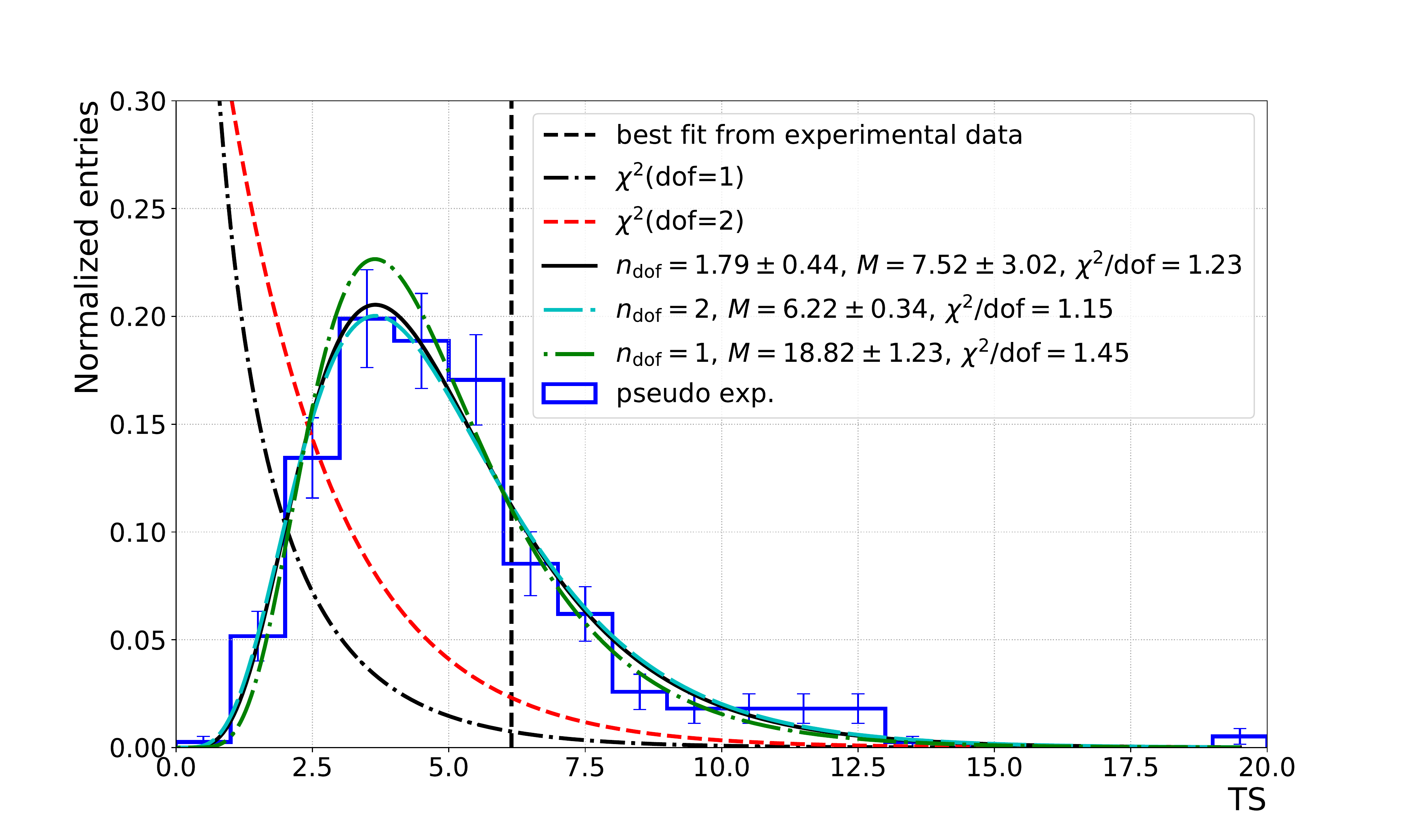}
    \caption{Expected distribution of test statistic values as obtained from \num{390} pseudo  experiments of the null hypothesis. Also shown is the expectation for a $\chi^2$-distribution with one and two degrees of freedom and various modified distribution functions (see text).}
    \label{fig:test-stat-global}
\end{figure}

As a consequence, the distribution of the test statistic values $TS =\lambda (\vec{x},  \vec{\eta}_0 )$ cannot be approximated by a $\chi^2 $-distribution but 
has to be derived from an ensemble study of pseudo experiments.
Due to the huge computational effort for scanning the full parameter space, this has been possible only for limited statistics of a few hundred pseudo experiments. The resulting test statistic values $TS$ when comparing the global minimum to the null hypothesis are shown in Fig.~\ref{fig:test-stat-global}.
It can be clearly seen that the test statistic strongly deviates from  $\chi^2 $-distributions
of one and two degrees of freedom.
Motivated by the fact, that the choice of the best of several random minima in the parameter space introduces a selection with trials  (often called  look-elsewhere effect), we introduce a trial factor in three versions of a modified approximation of the test statistic.
For this, we calculate the probability distribution $f_M(x) $
of the largest $\chi^2 $ value $x$ from an ensemble $M$ trials. This results to $f_M(x)
= M\cdot \chi^2(x,n_{dof}) \cdot \left ( \int_0^x \chi^2(y,n_{dof})\, dy  \right )^{M-1} $ where 
$\chi^2(x,n_{dof}) $ is the p.d.f. of a single trial.
Three versions of this approximation 
with $M$ as a free parameter are 
fitted to the pseudo experiments, 
using $\chi^2 $ distributions of one,  two, and a fitted degree of freedom $n_{dof} $.
All three versions  describe the observed test statistic reasonably well. Particularly the case of $n_{dof} = 2$
results in a fitted $M\simeq 6 $ which agrees well with the observations in pseudo experiments.

\begin{figure}[htbp]
    \centering
    \includegraphics[width=\columnwidth]{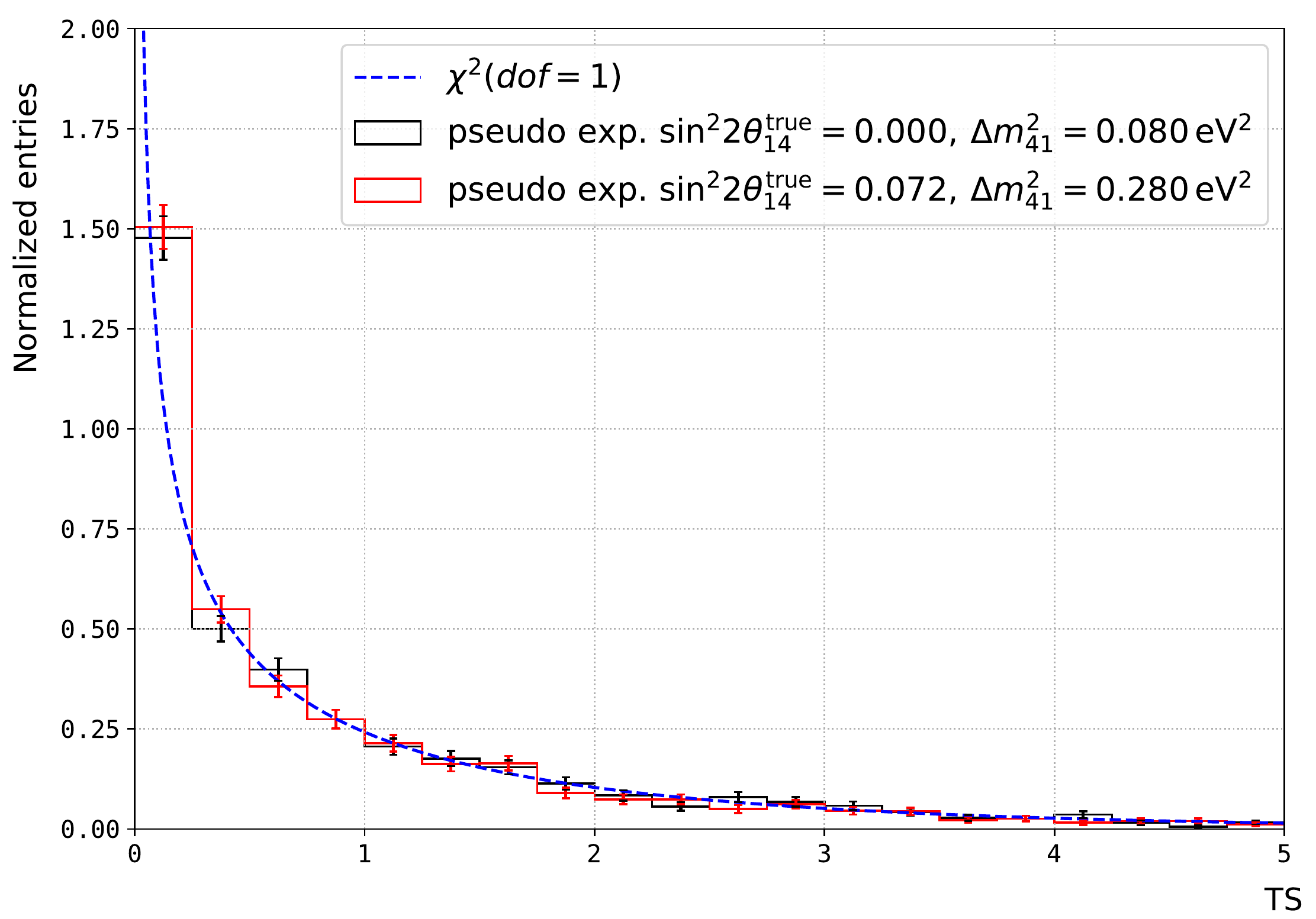}
    \caption{Test statistic for fixed values of  $\Delta m^2_{41}$. Shown are the results from \num{1999} pseudo experiments when fitting $\sin^2 2\theta_{14} $ for \num{100} discrete values of $\Delta m^2_{41}$ for the injected null hypothesis.  For comparison, the expectation from a $\chi^2$ distribution of one degree of freedom is shown. Additionally the test statistic for \num{1997} pseudo experiments of an injected signal is shown. Here the median of the fit $\sin^2 2\theta_{14}^{med} $ has been determined for each of the \num{100} tested  $\Delta m^2_{41}$ values. The test statistic is then evaluated with
     $\sin^2 2\theta_{14}^{med} $ instead of  $\sin^2 2\theta_{14} = 0 $ as the null hypothesis. All distributions are found to be consistent.
    }
    \label{fig:test-stat-local}
\end{figure}

The situation becomes simpler, when
taking into account that 
multiple values of $\Delta m^2_{41}$  can cause a minimum in the test statistic. In a modified hypothesis, we can define the sensitivity as the ability to test
values of $\sin^2 2\theta_{14} $ as a function of $\Delta m^2_{41}$. When analyzing the pseudo experiments  in a raster scan for distinct fixed values of  $\Delta m^2_{41}$ and varying only 
$\sin^2 2\theta_{14} $ \cite{Feldman:1997qc},
a distribution that is well compatible with the expectation
from a $\chi^2$ distribution with one degree of freedom is found as shown in Fig.~\ref{fig:test-stat-local}. 
Also for an injected signal, the test statistic with respect to the median expectation of the null hypothesis is consistent and also described by the same $\chi^2$ distribution.
This is a good confirmation of our assumption
that the observed trials are only related to 
different degenerated oscillation lengths.
This test shows that in this case the test statistic
can be well described 
with a $\chi^2$-distribution of one degree of freedom in agreement with  Wilks' theorem.

\section{Sensitivity}
We define the sensitivity,
in the following denoted as  Asimov-Wilks' (AW) sensitivity, by the boundary value $\sin^2 2\theta_{14} $  as a function of  $\Delta m^2_{41}$
 where the test statistic of the Asimov data set has a value $\langle \lambda (\vec{x}) \rangle \ge 3.84 $
(or $\langle \lambda (\vec{x} ) \rangle \ge 1 $).
This corresponds to the boundary of the median signal expectation where in case of  absence of a signal \SI{95}{\percent} (or \SI{68}{\percent})  of experiments obtain a smaller value of $\sin^2 2\theta_{14} $.
Note, that because an Asimov data set of the null hypothesis contains no fluctuations,
the use of Wilks' theorem is valid and not in contradiction with the above discussion.
Furthermore, the best found likelihood always corresponds to the injected null hypothesis. As the null hypothesis is degenerate in  $\Delta m^2_{41}$, this choice of sensitivity corresponds effectively  to a one-dimensional sensitivity on the maximum allowed value of $\sin^2 2\theta_{14} $ as a function of $\Delta m^2_{41}$, also known under the term raster-scan \cite{Lyons:2014kta}. This choice is consistent with the final choice of experimental limit that will be discussed below. Also, choosing a  $\chi^2$ distribution of one degree of freedom
for the test statistic value of \SI{95}{\percent} coverage is a result of this degeneracy.

This choice of sensitivity marks the region, where larger values
of  $\sin^2 2\theta_{14} $ are expected to lead to indications of a signal on the level of two (or one) standard deviations but is also closely related to the ability of constraining $\sin^2 2\theta_{14} $
in the absence of a signal.
These sensitivities are shown as lines in Fig.~\ref{fig:teststatistics} and Fig.~\ref{fig:toyexperiment}. The statistical coverage of the AW-sensitivity as well as the 
 unbiased estimation of the model parameters
$\sin^2 2\theta_{14} $ and  $\Delta m^2_{41}$ 
have been verified with ensembles of pseudo data in \cite{DeniseHellwig_PHD}.

For small values of  $\Delta m^2_{41} \lesssim \SI{5e-3}{eV^2}$, the sensitivity becomes weaker
as the disappearance becomes ambiguous with  conventional oscillations whose energy dependence is given by 
$\Delta m^2_{ee} $. The free nuisance parameter $\sin^2 2\theta_{13}$
becomes degenerate with $ \sin^2 2\theta_{14} $ and the sensitivity decreases.
Also towards large values of 
$\Delta m^2_{41} \gtrsim \SI{0.3}{eV^2} $ the sensitivity decreases, because oscillations become fast, and the disappearance turns into an
overall deficit 
for both detectors.
For the data-data fit approach as implemented here, an oscillation signal would thus become increasingly indistinguishable from an overall change of the reactor flux normalization. We have tested that by additionally constraining the  fit with a flux prediction. The sensitivity above 
$\Delta m^2_{41} \gtrsim \SI{0.3}{eV^2} $ would strongly improve but also become strongly model dependent. An interesting observation is the dip in sensitivity at $ \Delta m^2_{41} \simeq \SI{5e-2}{eV^2} $. The effect is related to the interference of maximum and minimum disappearance for neutrinos from the two reactor cores to the two detectors, whose baselines differ by 
about \SI{\sim100}{m}. A strong disappearance for signals of one of the reactor is counteracted by no disappearance for the other reactor.
We have tested that the effect disappears when simulating the baseline of only one reactor core.


\begin{figure}[htbp]
    \centering
 \includegraphics[width=\columnwidth]{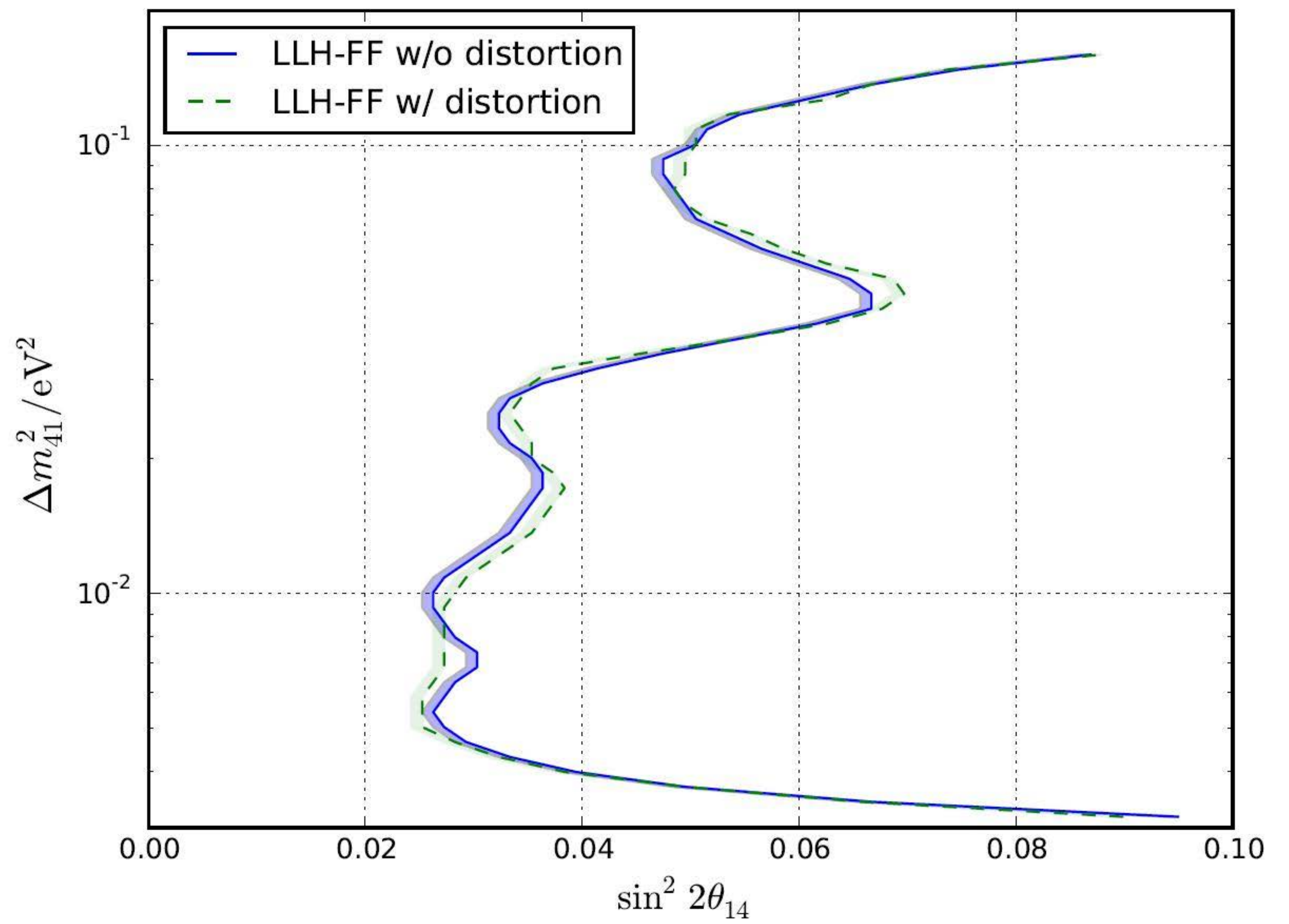}
    
    \caption{Sensitivity (95\% C.L.) of the analysis as obtained from Asimov data sets with and without a spectral distortion at \SI{5}{MeV}.}
    \label{fig:sensitivity-dist}
\end{figure}

The effect of the aforementioned spectral distortions of reactor flux models  has been studied with two Asimov data sets. One of them  included a  bump-like distortion at \SI{5}{MeV} using a double-Gaussian approximation of the measurement in \cite{An:2016srz}.
The resulting sensitivity is  only marginally impacted as shown in Fig.~\ref{fig:sensitivity-dist}.

\section{Experimental result}

\begin{figure}[htbp]
    \centering
    \includegraphics[width=\columnwidth]{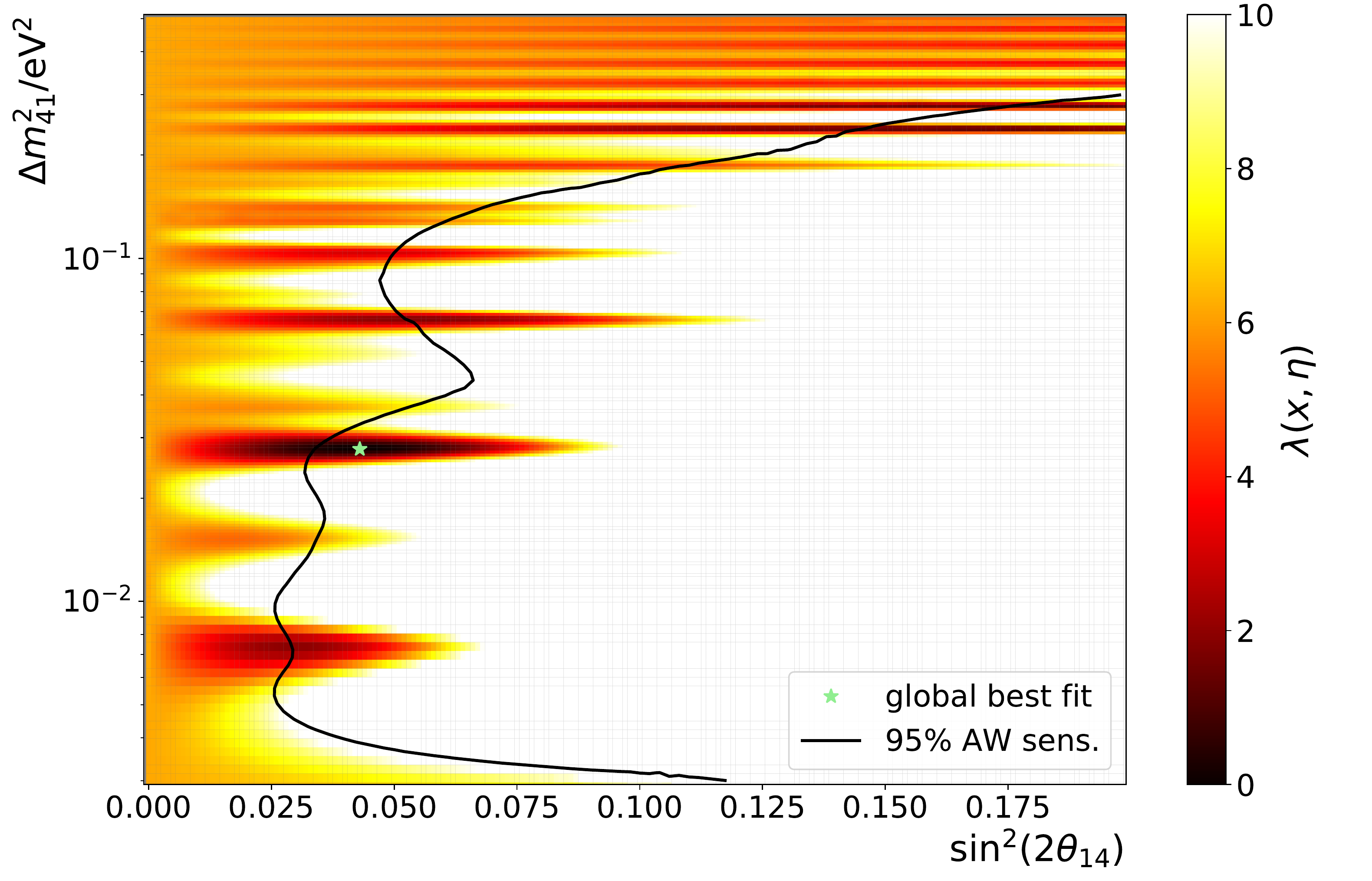}        \caption{Likelihood scan of the experimental data.}
    \label{fig:data-contour}
\end{figure}

The result of the scan of the test statistic $\lambda(\vec{x} ) $ for the experimental data is shown in Fig.~\ref{fig:data-contour}. 

The global best fit minimum is found for the values 
$ \hat{\sin^2 2\theta_{14}}  = 0.043 $ and 
$ \hat{\Delta m^2_{41}} = \SI{0.028}{eV^2}$.
The  nuisance parameters converged to values within their reasonable range. In particular the best fit value 
$\sin^2 2\theta_{13} =
{0.108^{+0.016}_{-0.017}}$ 
of the null hypothesis
is found in agreement 
with the nominal value 
\num{0.105+-0.014} 
that has been obtained from the same data set \cite{DoubleChooz:2019qbj}.
The difference to that result is  expected from
the  differences of the fit method and has been 
verified in a  detailed comparison of the fit methods.
Also the value $ \sin^2 2\theta_{13} = 0.1077 $ obtained for the global best fit $ \hat{\eta} $ is very close to the null hypothesis and thus does not indicate a pull on the best fit.

The value of the test statistic of the best fit with respect to the null hypothesis of no sterile mixing is $\lambda(\vec{x}_{exp} ) = 6.15 $.
From \num{388} performed pseudo experiments of the null hypothesis in Fig.~\ref{fig:test-stat-global},
a total of \num{96} have a larger or equal value of $\lambda$. The corresponding p-value is \SI{24.7 \pm 2.2 }{\percent}.
This p-value does not depend on details of the modeling of the test statistic. When using the three approximations of the test statistic distributions 
in Fig.~\ref{fig:test-stat-global},  very similar p-values
between  \SIrange{22}{26}{\percent} are obtained.
Therefore, 
the experimental result is fully consistent with the null hypothesis of no mixing with sterile neutrinos and no evidence for a signal can be reported.

The location of the best-fit point is not within the region of good sensitivity but  close to the estimated sensitivity line, see Figs.~\ref{fig:sensitivity-dist} and \ref{fig:teststatistics}. This is, as  discussed above,  an expected feature of statistical background fluctuations that are being
picked up by a signal model.

\begin{figure}[htbp]
    \centering
    \includegraphics[width=\columnwidth]{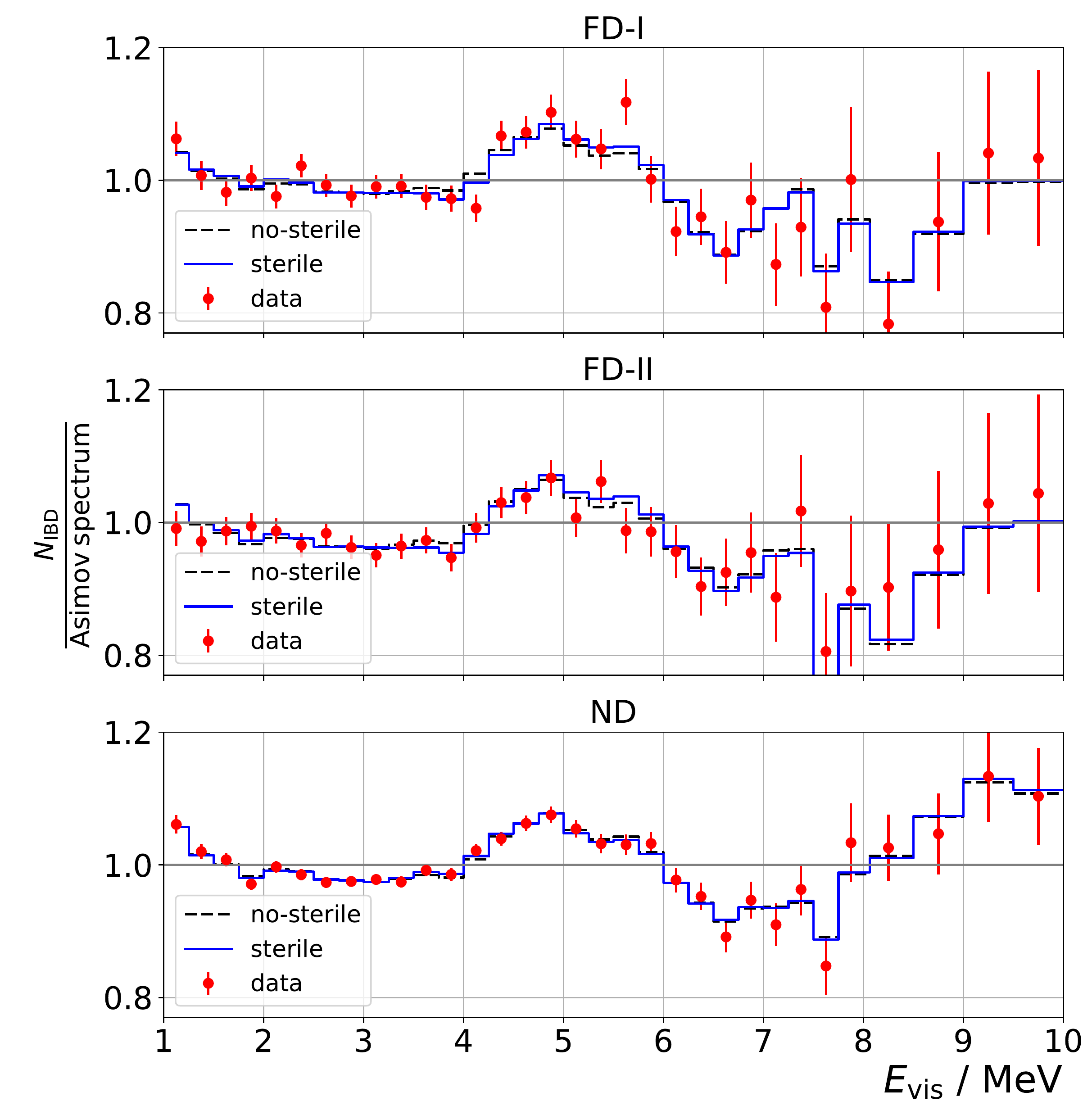}
    \caption{Experimental residuals for the three consistently fitted data sets FD-I, FD-II, and ND. The data are normalized
    to the nominal reactor expectation \cite{Huber:2011wv} adapted to the Double Chooz reactors including conventional oscillations with parameters taken from an independent measurement \cite{Adey:2018zwh}. The experimental data are plotted as red dots. The global best fit is shown as a solid line while for comparison the best-fit null hypothesis is shown as a dashed line.
    As the fit optimizes systematic uncertainties to the data, only statistical error bars are displayed.
    }
    \label{fig:residuals}
\end{figure}

Figure \ref{fig:residuals} shows the fit residuals normalized to the number of events expected for the nominal reactor-model including conventional oscillations. Also shown are the best fit of the null (non-sterile) and best-fit sterile hypothesis. 
All three data sets are consistently  described by both models with a generally good agreement,
including the observed bump at \SI{5}{MeV} and other spectral features, as expected from the  implementation of the fit. 
No particular difference is observed between the three data sets that would hint to a mismodeled 
detector responses.
Note that due to the use of a free
but global normalization for each energy bin, the fit does not  depend on the assumed shape and normalization of the initial reactor flux model but only on the consistency of the measured experimental data in the three data sets. The sterile model achieves a marginally better description. The difference can be quantified by  Pearson's $\chi^2 $-test 
\cite{Tanabashi:2018oca}. 
The summed  $\chi^2$ values of the three data distributions of Fig. \ref{fig:residuals} are \num{78.17} for the best-fit no-sterile model and \num{71,91} for the best-fit sterile model, respectively. With a rough estimation of the number of degrees of freedom of \num{76}, i.e. the number of data points corrected for the free overall normalizations of each energy bin, this indicates an acceptable goodness of fit for both models.
The difference  $\Delta \chi^2  = 6.25$ shows no systematic trend and
is largely driven by a few fluctuating data points, i.e.\ the two energies  \SIlist{4.1;5.6}{MeV} dominate the difference with a summed contribution of $\Delta \chi^2 = 5.5$.
As discussed above, this is an expected behavior also for the no-sterile case where for each statistical fluctuation of data a matching sterile hypothesis can be constructed. No general trend in the data supporting a sterile signal is  observed, which is consistent with the observation of an insignificant p-value as reported above.

\section{Discussion}

The experimental data has been tested over the full range of the two-dimensional signal parameter space.
The globally found minimum 
does  not constitute a significant observation of a signal but is well compatible with the null hypothesis of no mixing with sterile neutrinos.

In response to the 
limited computing resources that do not permit the evaluation of the test statistic with pseudo experiments at every point in the two-dimensional parameter space with sufficiently  accurate coverage, we have decided for a more robust limit-setting strategy which is also known under the term \emph{raster-scan} \cite{Lyons:2014kta}.
Here, we calculate one-dimensional exclusion limits on the maximum allowed value of  $\sin^2 2\theta_{14} $ as a function of
  $\Delta m^2_{41}$.
  These limits are calculated with a frequentist approach based on Wilks' theorem
  comparing  the local test statistic with respect to the best fit at the probed $\Delta m^2_{41}$ and using the $\chi^2$ probability with one degree of freedom.
    The statistical coverage of the approach has been verified with pseudo experiments of injected signal as shown above.
    
    Alternatively a two dimensional approach
    could be pursued, where the test statistic is compared to the globally found maximum likelihood. Such a strategy has  been followed e.g.\ for the analysis in Daya Bay \cite{An:2016srz}. Here the exclusion would correspond
    to the probability of the combination of 
    $\sin^2 2\theta_{14} $ and
  $\Delta m^2_{41}$. However, pseudo experiments with an injected signal have revealed
  that our test statistic strongly depends on the injected value of  $\sin^2 2\theta_{14} $.
  For small values of  $\sin^2 2\theta_{14} $ it is close to the test statistic that we have observed for the null hypothesis (see Fig.~\ref{fig:teststatistics}) while it  gradually crosses over into  a $\chi^2$-distribution of two degrees of freedom for larger values. Because the determination of limits with correct statistical coverage would require the simulation of a very large number of pseudo experiments throughout the entire parameter space, we have chosen   the raster-scan approach.
Within the available computing resources this resulted in limits of more accurate coverage.
  We note that this strategy  applies only to the setting of limits but not to the p-value of the analysis that has been obtained in a full two-dimensional approach.

\begin{figure}[htbp]
    \centering

    \includegraphics[width=\columnwidth]{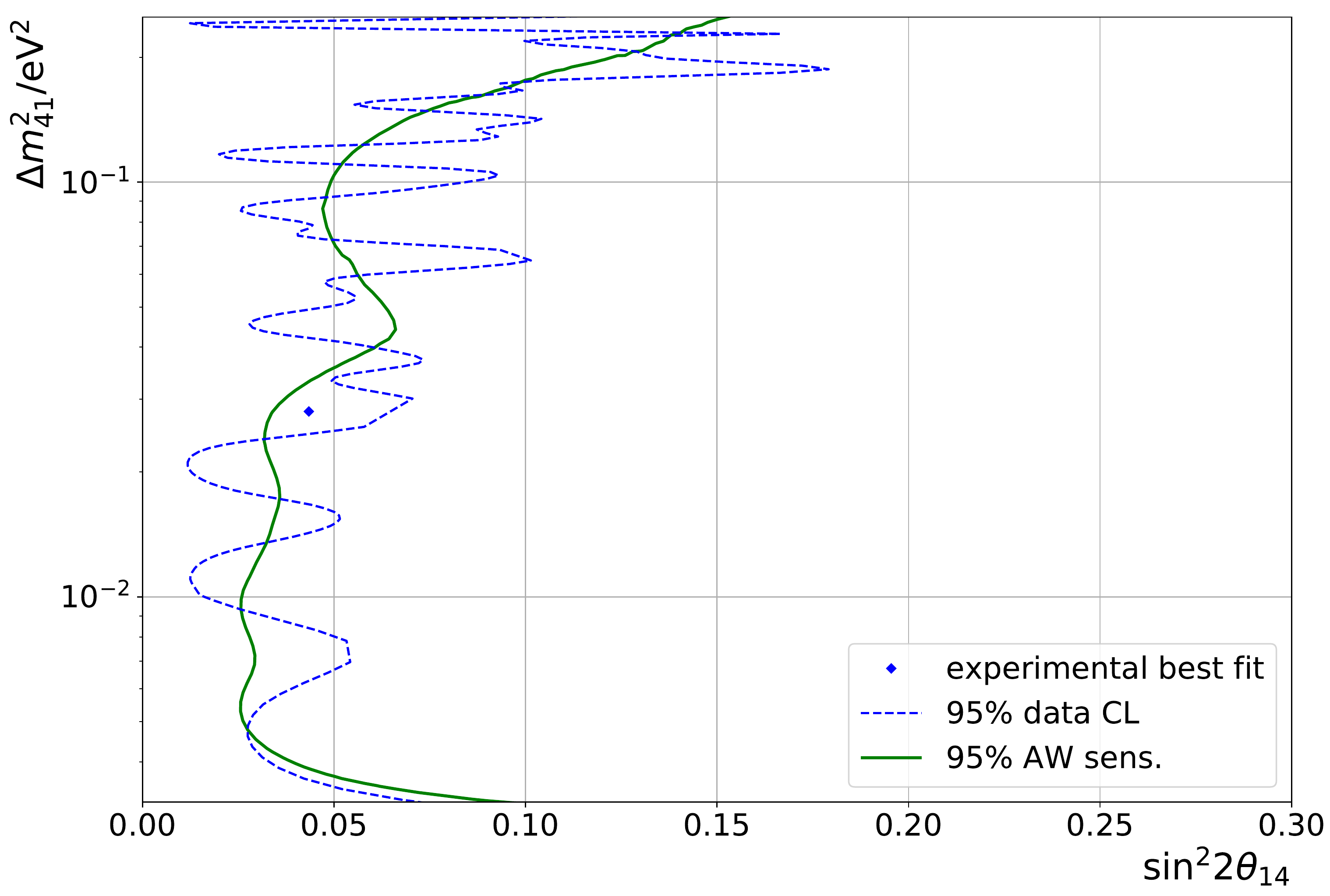}
    \caption{Upper limit (dotted blue line) at \SI{95}{\percent} C.L.\ for $\sin^2 2\theta_{14} $ as a function of
  $\Delta m^2_{41}$. The black dot indicates the position of the global best fit.
    The green solid  line corresponds to the 
    AW-sensitivity.   
   }
    \label{fig:exclusion-contour}
\end{figure}

The resulting exclusion limits are shown in Fig.~\ref{fig:exclusion-contour}. The obtained limits are generally close to the AW-sensitivity.
For masses $\Delta m^2_{41}$, where the best fit results in the null hypothesis $\sin^2 2\theta_{14} =0 $, the upper limit coincides  with
the median expected sensitivity.
Due to statistical fluctuations in the data one expects variations around this median sensitivty, 
depending whether excesses or deficits in the prediction match  these fluctuations better.
 As the allowed parameter space is bounded to positive values of  $\sin^2 2\theta_{14}  $, we  expect  roughly  for \SI{50}{\percent}
 of probed $\Delta m^2_{41}$ values fits with a non-zero  value of $\sin^2 2\theta_{14} $ resulting in less constraining limits than the average sensitivity and similarly a roughly equal number of more constraining limits.

\begin{figure}[htbp]
    \centering
    \includegraphics[width=\columnwidth]{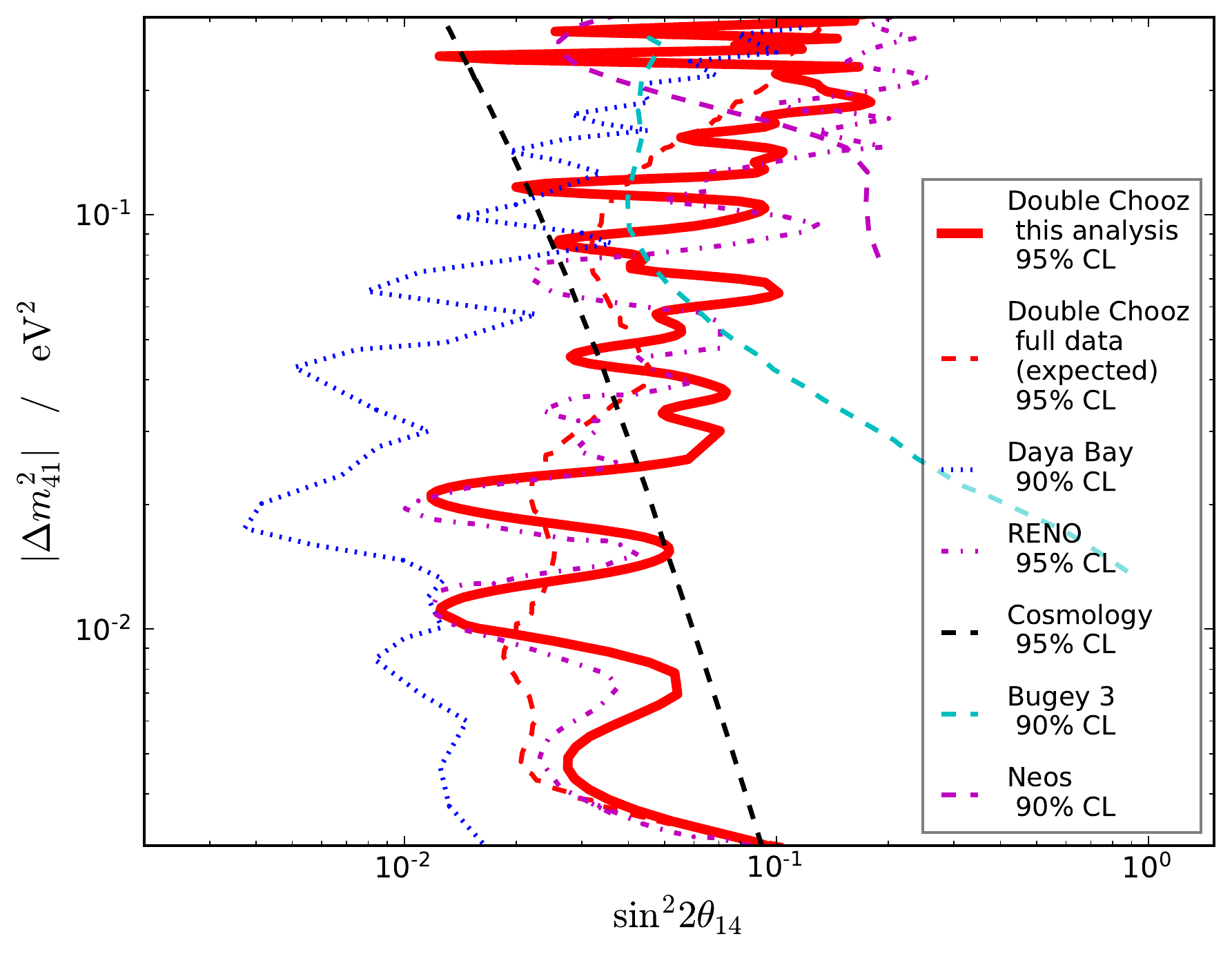}
    \caption{Comparison of the upper exclusion limits provided by this analysis (Double Chooz) with results from other measurements:  Daya Bay \cite{Adamson:2020jvo}, RENO \cite{Choi:2020ttv}, 
    Bugey-3 \cite{Declais:1994su} Neos \cite{Ko:2016owz}, and cosmological limits \cite{Adams:2020nue} based on the combination of observations of the cosmic microwave background, gravitational lensing and baryon-acoustic oscillations. Additionally displayed is the expected average sensitivity of Double Chooz with the full data statistics from the multi-detector phase. Note, that the figure combines 1-d and 2-d limits as well as limits of different confidence level.}
    \label{fig:result-compare}
\end{figure}

As discussed above, both experiments Daya Bay and RENO probe a similar range of $L/E $ values and have published exclusion limits for a similar range of $\Delta m^2_{41}$ for sterile neutrino 
mixing in the 3+1 model \cite{Adamson:2020jvo,An:2016luf,Choi:2020ttv}.
A comparison of these results is shown in Fig.~\ref{fig:result-compare}. We note, that a detailed quantitative comparison is difficult, because, unlike Double Chooz and as discussed above, the two other experiments provide two-dimensional limits. In order to quantify the difference, 
we have evaluated  the consistency of the two-dimensional and one-dimensional raster-scan approach with seven pseudo experiments of the null-hypothesis. It is found that the median line
of the 2d TS value of 8.7, corresponding to a
\SI{10}{\percent} p-value in figure \ref{fig:test-stat-global} matches within about \SIrange{10}{20}{\percent} with the \SI{90}{\percent} AW-sensitivity.
Furthermore, marginalizing the TS values of the 2-d contours of these pseudo experiments for every $\Delta m_{41}^2 $ value and averaging the pseudo experiments
matches well with the 1d AW-sensitivity  without noticeable bias.
Another important difference with respect to the aforementioned experiments is that analysis assumptions differ. In particular, the Daya Bay  result includes a reactor flux model and constraints on $\theta_{13}$. 
We have tested that such assumptions would also increase the sensitivity of this analysis. The statistics of $\nubar_e $ candidates used in Daya Bay and RENO is roughly four times the statistics used here.
In addition, the figure shows limits obtained by the Bugey-3 
collaboration and limits from combining cosmological observations.
The Double Chooz result based on the here used data is less constraining than Daya Bay but is competitive to the other presented results.

The result has been obtained under the assumption of a 3+1 model. An extension to a 3+2 model would require the extension of the $3\times3$  PMNS Matrix to $5\time 5$ dimensions
with  \num{7}  additional mixing angles 
plus additional CP phases and the oscillations
would also involve additional mass differences.
In the simplest approximation, equation (\ref{eq:oscillation}) would include an additional term $-\sin^2 2\theta_{15} \sin^2\Delta m^2_{51} L/(4E) $. This leads to additional oscillations, which potentially interfere with the 4-1 oscillation if $\Delta m^2_{41}  \approx \Delta m^2_{51} $. As a result of test studies \cite{DeniseHellwig_PHD}, 
we find that the here presented limits of the mixing angle as a function of $\Delta m^2 $ are largely valid also for  3+2 models with largely different mass difference and in particular if $\Delta m^2_{51} \gtrsim \SI{0.3}{eV^2} $. In case both mass-square differences fall into the sensitive region of this analysis, the oscillation of the respective larger $\Delta m^2 $ is largely washed out and results in a global normalization offset, to which the data-to-data fit of this analysis is insensitive. In summary, though different in statistical coverage, the test for a 3+1 model is also sensitive for a signal of a more complicated model.

The relative impact of systematic uncertainties has been 
tested in terms of sensitivity for the null hypothesis and for relatively strong signals of $\sin^2 2\theta_{14} =0.1$ and varying values of $\Delta m^2_{41}$. It is found that the relative impact of systematic uncertainties on the total error increases towards smaller values  $\Delta m^2_{41}$. E.g.\ for determining the value  $\sin^2 2\theta_{14} =0.1$  the relative error changes  
from ${\sigma_{stat} \over \sigma_{tot} } = \SI{99}{\percent}$ for $\Delta m^2_{41} =
\SI{0.1}{eV^2}$ to  ${\sigma_{stat} \over \sigma_{tot} } = \SI{55}{\percent}$ for $\Delta m^2_{41} = \SI{7.3e-3}{eV^2}$.
Among the different systematic parameters, the uncertainty of the energy scale and the unconstrained parameter $\theta_{13}$ show the largest impact on the total uncertainty.
As the current analysis is 
limited by statistics, it will benefit from 
the  full  data set of Double Chooz. Figure \ref{fig:result-compare} shows the expected
median sensitivity for the full duration of multiple detector operation, corresponding to
an increase in statistics by roughly a factor \num{2.4}.
In addition, we expect improvements by 
the off-reactor data set, that is enlarged from 7 to 32 days, resulting in 
reduced uncertainties in background modeling and furthermore the planned improved measurement of the proton number of the neutrino target.

\section{Summary}

We have presented  an initial  search  for  oscillations of electron anti-neutrinos with additional sterile neutrino flavors with the Double Chooz experiment. The search uses data from five years of operation of Double Chooz, including  two  years  in  the  two-detector  configuration.  The analysis method is based on a profile likelihood, searching for the disappearance due to oscillations in a data-to-data comparison of the two respective detectors.  The  analysis  is  optimized  for  a  3+1  model  and  is sensitive in the  mass range $\SI{5e-2}{eV^2} 
\lesssim \Delta m^2_{41} \lesssim \SI{3e-1}{eV^2} $.
No significant disappearance signal additionally to the conventional  oscillations related  to $\theta_{13} $ is  observed  in a full 2-d scan of the model parameters. Correspondingly exclusion bounds on the sterile mixing parameter
are determined in form of a raster-scan of
$\theta_{14}  $ as a function  of  $ \Delta m^2_{41} $.
The result is competitive to similar searches in this mass range.
An update to the full data set from Double Chooz is planned.

\begin{acknowledgements}

We thank the EDF ("Electricity of France") company; the European
fund FEDER; the Région Grand Est (formerly known as
the Région Champagne-Ardenne); the Département des Ardennes;
and the Communauté de Communes Ardenne Rives de Meuse.
We acknowledge the support of the CEA, CNRS/IN2P3, the computer
centre CC-IN2P3 and LabEx UnivEarthS in France; the Max
Planck Gesellschaft, the Deutsche Forschungsgemeinschaft DFG,
the Transregional Collaborative Research Center TR27, the excellence
cluster "Origin and Structure of the Universe" and the
Maier-Leibnitz-Laboratorium Garching in Germany; the Ministry
of Education, Culture, Sports, Science and Technology of Japan
(MEXT) and the Japan Society for the Promotion of Science
(JSPS) in Japan; the Ministerio de Economía, Industria y Competitividad
(SEIDI-MINECO) under grants FPA2016-77347-C2-1-
P and MdM-2015-0509 in Spain; the Department of Energy and
the National Science Foundation; the Russian Academy of Science,
the Kurchatov Institute and the Russian Foundation for Basic
Research (RFBR) in Russia; the Brazilian Ministry of Science,
Technology and Innovation (MCTI), the Financiadora de Estudos
e Projetos (FINEP), the Conselho Nacional de Desenvolvimento
Científíco e Tecnológico (CNPq), the S\~{a}o Paulo Research Foundation
(FAPESP) and the Brazilian Network for High Energy Physics
(RENAFAE) in Brazil.

\end{acknowledgements}

\bibliographystyle{spbasic}      
\bibliography{dc}   

\end{document}